\documentclass[preprint,journal]{vgtc}            

\onlineid{0}

\ieeedoi{10.1109/TVCG.2024.3456394}

\vgtccategory{Research}

\vgtcpapertype{please specify}

\title{2D Embeddings of Multi-dimensional Partitionings}

\author{%
  \authororcid{Marina Evers}{0000-0003-3904-5065} and \authororcid{Lars Linsen}{0000-0002-6168-8748}
}

\authorfooter{
  \item
  	All authors are with the University of Münster, Germany. \\
  	E-mail: \{marina.evers\textbar linsen\}@uni-muenster.de.
}

\abstract{%
Partitionings (or segmentations) divide a given domain into disjoint connected regions whose union forms again the entire domain. Multi-dimensional partitionings occur, for example, when analyzing parameter spaces of simulation models, where each segment of the partitioning represents a region of similar model behavior. Having computed a partitioning, one is commonly interested in understanding how large the segments are and which segments lie next to each other. While visual representations of 2D domain partitionings that reveal sizes and neighborhoods are straightforward, this is no longer the case when considering multi-dimensional domains of three or more dimensions. We propose an algorithm for computing 2D embeddings of multi-dimensional partitionings. The embedding shall have the following properties: It shall maintain the topology of the partitioning and optimize the area sizes and joint boundary lengths of the embedded segments to match the respective sizes and lengths in the multi-dimensional domain. We demonstrate the effectiveness of our approach by applying it to different use cases, including the visual exploration of 3D spatial domain segmentations and multi-dimensional parameter space partitionings of simulation ensembles. We numerically evaluate our algorithm with respect to how well sizes and lengths are preserved depending on the dimensionality of the domain and the number of segments. 
}


\teaser{
  \centering
  \includegraphics[width=\linewidth, alt={A cube divided in 8 parts with colors shows the multi-dimensional partitioning. The second step shows graph embedding with straight lines where nodes and links have the same colors as the partitions in the cube. The third step shows a rectangle that is filled with segments in the corresponding colors. Edge crossings are indicated by black pixels, and background pixels remain gray. The fourth step shows the rendering. At the borders of the segments, the surfaces appear to decrease.}]{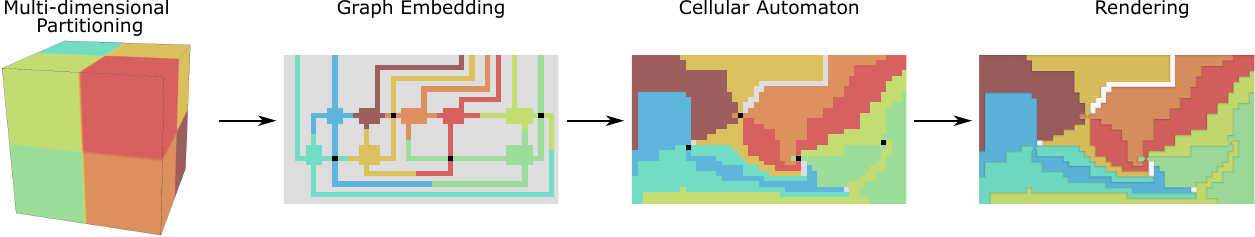}
  \caption{%
  	A multi-dimensional partitioning is modeled as a graph that is embedded into a 2D plane. The graph embedding is used as a starting point for computing an area- and boundary-length-preserving layout of the partitioning using a cellular automaton approach. To its outcome, we apply a rendering that highlights relevant features.%
  }
  \label{fig:overview}
}

\graphicspath{{figs/}{figures/}{pictures/}{images/}{./}} 

\usepackage{tabu}                      
\usepackage{booktabs}                  
\usepackage{lipsum}                    
\usepackage{mwe}                       
\usepackage{amsmath}

\usepackage{mathptmx}                  

\newenvironment{denseitemize}%
{\begin{itemize}\setlength{\itemsep}{-3pt}}%
{\end{itemize}}

\begin{document}


\firstsection{Introduction}

\maketitle

Given a multi-dimensional domain, partitionings or segmentations split the domain into connected regions. The regions are disjoint, yet neighboring regions share a boundary. The union of the regions represents the entire domain again. Such multi-dimensional partitionings occur, for example, in parameter space analysis of simulation ensembles. Here, a common analysis goal is to identify regions in parameter space that share a common behavior in the simulation outcome. Especially for parameter spaces with more than three dimensions, it becomes challenging to understand the structure of the partitioning. Besides investigating the number of segments, it is of interest to understand how large the parameter-space region for each segment is, as this property provides information about the stability of the simulation output with respect to the input parameters. Additionally, the neighborhood relations between different segments provide information about possible transitions between different behaviors. Thus, a visualization of the partitioning should display all segments simultaneously, should reveal neighborhoods, and should visualize segment sizes and boundary lengths of neighboring segments to cover the most important structural information of the partitioning.

When analyzing image segmentations, the segmentation outputs are commonly visualized by assigning different colors to different segments, which works well for 2D domains, but not anymore for higher-dimensional domains, not even for 3D segmentations. Despite the fact that 3D images are commonly acquired and segmented in different fields such as medicine, geology, or material science, there exists no approach, to our knowledge, that globally visualizes the segmentation output similar to 2D domains. 

We present an approach for visualizing a multi-dimensional partitioning or segmentation in a 2D embedding. Similar to 2D segmentation visualizations, we encode segments in the 2D embedding by color. The embedding represents all segments, preserves the topology of the multi-dimensional segmentation, and optimizes for the preservation of segment sizes and boundary lengths. The computation of the embedding is based on a graph representation of the multi-dimensional segmentation and a respective graph embedding. The optimization is based on a cellular automaton approach. The output of the optimization is visualized by a shading technique for boundaries and a visual encoding of edge crossings, which cannot be avoided in case of non-planar graphs.

Our main contributions can be summarized by:
\begin{denseitemize}
    \item An algorithm for computing a 2D embedding of a multi-dimensional segmentation that preserves the topology of the given multi-dimensional segmentation.
    \item An optimization of the embedding to generate segment areas and boundary lengths that match segment sizes and boundary sizes of the multi-dimensional segmentation.
    \item A rendering to highlight features of the embedding such as segment boundaries and edge crossings.
    \item Application of our approach to visualizing 3D spatial domain segmentations and $n$-dimensional parameter space partitionings.
    \item An in-depth evaluation of the algorithm's performance based on its optimization criteria. 
\end{denseitemize}

\section{Related Work}
\textbf{Segmentation of 3D images} are ubiquitous in medical image analysis, among others. Visualizing the segmentation output is typically performed by showing 2D slices of the volume or focusing on individual segments of interest~\cite{preim2007visualization}, i.e., not providing a global representation of the segmentation output. Similarly, in volume renderings, transfer functions are applied to assign colors and opacities to regions of interest to alleviate the inherent occlusion problem, but also do not provide a visual representation of the entire 3D segmentation. 

Multi-dimensional spaces beyond 3D occur in simulation ensemble analysis~\cite{Wang2019, Crossno2018} when studying the parameter space of the ensembles. \textbf{Parameter space partitionings} represent the different behaviors that the simulated phenomenon may have. Sedlmair et al.~\cite{sedlmair2014visual} present a framework for visual parameter space analysis that identifies partitionings as one of the core analysis tasks. Paraglide~\cite{bergner2013paraglide} provides an interactive visual analysis of parameter space partitionings using scatterplots, but does not focus on understanding the partitioning's structure in the multi-dimensional space. Other works~\cite{Orban2019, spence1995visualization} use projections of parameter spaces, but operate on individual points instead of partitionings.
Further techniques for the visualization of parameter spaces are radial layouts~\cite{bruckner2010result} and glyph layouts~\cite{bock2015visual}, which do not explicitly show the partitioning with its topology. This also holds for parallel coordinates, which are a common tool in visualizing multi-dimensional data and have been adapted for investigating parameter spaces~\cite{obermaier2015visual, Wang2017}. Closest to our work are the following two approaches for direct visualization of partitionings: Evers et al.~\cite{evers2022multi} use an enriched hyper-slicer for a distortion-free visualization of multi-dimensional parameter space partitionings. However, their approach does not provide a high-level overview of all segments including topology and segment sizes. Fernandes et al.~\cite{fernandes2019visual} study region transitions in partitionings by a glyph-based approach. They include areas of the segments and boundary sizes between them, but as the glyphs are relatively complex, the approach does not scale well to a higher number of segments.
However, none of these approaches provides an overview of the multi-dimensional partitioning.

Our approach uses a graph representation of the multi-dimensional segmentation. Deriving a \textbf{graph from a segmentation} has been proposed by Ren et al.~\cite{ren2017joint} to visualize joint layouts for segmented meshes. However, they do not aim at visualizing the properties of the segments. After deriving the graph from a segmentation, the graph needs to be visualized while preserving the segmentation's properties. Various graph drawing techniques with different characteristics have been discussed~\cite{tamassia2013handbook, von2011visual, mcgee2019state, vehlow2017visualizing, fischer2021towards, didimo2019survey, schulz2006visualizing, herman2000graph}. Common techniques include force-directed layouts~\cite{kamada1989algorithm, fruchterman1991graph}, layered graph drawing~\cite{sugiyama1981methods} and orthogonal layouts~\cite{batini1986layout}.

For \textbf{graph visualization}, besides classical node-link diagrams, different space-filling visualization techniques have been proposed. Treemaps~\cite{shneiderman1992tree} visualize weighted trees and provide a good sense of size-related tasks~\cite{teoh2007study, fiedler2020survey}, but our segmentation does not resemble a tree structure. Also, techniques that focus on visualizing clusterings in graphs are related to our problem. They include BubbleSets~\cite{collins2009bubble}, LineSets~\cite{alper2011design}, and KelpFusion~\cite{meulemans2013kelpfusion}, which add regions around previously placed vertices. GMap~\cite{gansner2010gmap, hu2010visualizing} creates map-like visualizations of clustered graphs, but the layout might be highly fragmented. While this problem can be overcome by modifying the layout or the clustering~\cite{kobourov2014visualizing}, the resulting visualization does not ensure joint boundaries in the visualization in case of connections between the clusters. MapSets~\cite{efrat2014mapsets} targets a similar problem as GMap, but might lead to complex regional layouts. 

Preserving areas in a given topology is strongly related to \textbf{cartograms}~\cite{nusrat2016state}. 
Layouts where shared boundaries present neighborhoods have been proposed~\cite{alam2012computing, yeap1993floor}, but they only apply to planar graphs. Wu et al.~\cite{wu2020multi} propose a visualization method for clustered graphs that deal with vertices belonging to more than one cluster by drawing connections on top of the layout. However, these connections are structurally different from the visualization of the clusters. Thus, in our case, a similar approach to dealing with non-planar graphs is not desirable. 

\section{Overview}
\label{sec:overview}
We propose a visual encoding that represents a multi-dimensional segmentation and its most important features. As the driving application scenario for the development of our approach is the analysis of simulation ensembles, we derive tasks and corresponding optimization criteria based on analyzing multi-dimensional parameter spaces. 
While the goal of analyzing multi-dimensional parameter spaces motivates our design choices, the developed approach is nevertheless generally applicable to the visualization of any domain segmentation.

Motivated by the application to the analysis of multi-dimensional parameter spaces, our overarching goal is to obtain an overview of the segmentation, identified as a relevant task in previous work~\cite{evers2022multi}. To achieve this goal, we identified the following tasks:\\
\textbf{T1}: \textit{Identify neighborhood relations.} The topological structure of a segmentation provides important information. In particular, it reveals which transitions between parameter space segments are possible.\\
\textbf{T2}: \textit{Understand the relative sizes of the segments.} When analyzing a multi-dimensional parameter space, the users should be able to identify which portion of the parameter space corresponds to the respective segment, i.e., how likely the respective behavior is.\\
\textbf{T3}: \textit{Understand the relative sizes of the boundaries between segments.} The sizes of the boundaries indicate how large the parameter ranges are that allow for direct transitions between the two considered segments. Our visualization should enable users to estimate whether there is a strong connection between the segments, which corresponds to a large boundary size, or whether the segments are only slightly connected.

Based on these tasks, we can define the following properties for our visual design: \\
\textbf{Preserving Topology}. The topology of a segmentation is defined by neighborhood relationships between the segments, thus, preserving it addresses Task T1.\\
\textbf{Representing Segment Sizes}. The sizes of multi-dimensional segments might vary significantly. Therefore, the sizes should also be represented in the lower-dimensional embedding, addressing Task T2. 
\\
\textbf{Representing Segment Boundary Sizes}. To address Task T3, the embedding should be optimized for the boundary sizes. Note that in this paper we refer to the boundary sizes as boundary lengths to avoid confusion with the segment sizes, even though only in 2D they are actually lengths.

Note that we do not consider the shapes of the segments for our design, as it is infeasible to represent multi-dimensional shapes in a single 2D embedding.

\noindent
\textbf{Approach.}
We propose an embedding algorithm for visualizing a multi-dimensional segmentation in two dimensions. The algorithm consists of several steps shown schematically in \cref{fig:overview}. 
As input, we consider a segmented multi-dimensional volume of dimensionality $n$, where each segment represents a connected component and is labeled with a unique ID.
The segments partition the $n$-dimensional domain such that each point in the $n$-dimensional domain is assigned to exactly one segment. 
Without loss of generality, we assume that the domain is given in a discrete setting in the form of a regular grid, i.e.,  each grid point stores the assigned segment's ID.
This assumption is only made for implementation purposes.
If the multi-dimensional segmentation is given in another format, it can be easily resampled to a regular grid.

The computation of the embedding (see \cref{sec:embedding}) is based on a graph representation of the multi-dimensional segmentation (see \cref{sec:graph-computation}). 
The graph's vertices represent the segments, its edges represent neighborhood information,  segment sizes are stored as vertex weights, and boundary lengths are stored as edge weights. 
We, then, compute a 2D embedding of the graph that minimizes edge crossings. 
Afterwards, the graph embedding is used as input to a cellular automaton approach, which operates on a 2D grid of cells that is initialized with a drawing of the graph embedding and eventually labels each cell with a segment ID. The cellular automaton approach iteratively optimizes the area size of the segments and the boundary lengths between segments to match the segment sizes and boundary sizes of the multi-dimensional segmentation while preserving the topological structure (see \cref{sec:automaton}). We choose a cellular automaton over other optimization algorithms because it can guarantee topology preservation and can be applied to the discrete, non-differentiable segmentation data. Moreover, other algorithm choices that would be applicable, such as genetic algorithms or simulated annealing, are usually not more efficient.
Finally, we present a rendering approach for the output of the cellular automaton that highlights its important features (see \cref{sec:visualization}).

\section{Embedding Algorithm}
\label{sec:embedding}
\subsection{Graph Representation}
\label{sec:graph-computation}
The topological structure of the segmentation $S$ can be represented by a weighted, undirected graph $G=(V,E)$. Each segment $s_i \in S$ corresponds to a vertex $v_i \in V$, and a shared boundary of two segments $s_i, s_j \in S$ is represented by an edge $(v_i, v_j) \in E$. In addition, there are also segments that lie at the border of the given domain. This property should also be preserved. Therefore, we add an additional vertex $v_b$ representing the outer border of the domain. For each vertex $v_i$ representing a segment connected to this outer border, we add a new edge $(v_i, v_b)$ to the graph. For our implementation that assumes a segmentation over a regular grid, we assume two segments to be neighbors, if and only if two grid cells of the two segments share a common face along one of the dimensions, i.e. in 3D, cells that only share a vertex or an edge are not considered to be neighbors. 
In the context of cellular automata, this is referred to as von Neumann neighborhoods. For higher-dimensional data, the neighborhood structure of von Neumann neighbors is very sparse as only the neighbors along a single dimension are considered~\cite{zaitsev2017generalized}. For the applications presented in this work, this does not impose problems since the dimensionality of none of the presented examples is very high. Additionally, the neighborhood criterium for creating the graph and computing the boundary lengths can be easily exchanged without influencing the later stages of the algorithm. 

We further enhance our graph representation with weights for vertices and edges. The size $A_i$ of a segment $s_i$ is stored as a weight of vertex $v_i$, which later will be used to preserve the corresponding area in the 2D embedding. The size $A_i$ is computed by counting the number of $n$-dimensional cells that the segment consists of. 
The size of the boundary between two segments $s_i$ and $s_j$ is stored as a weight $B_{ij}$ of edge $(v_i, v_j)$ and is computed by counting the number of cells that are von Neumann neighbors. %

Given a graph $G$ that represents the multi-dimensional segmentation, we next want to embed this graph into two dimensions, where the embedding shall minimize edge crossings. Edge crossings make a graph layout more complex and, thus, harder to understand. Additionally, they represent a structure that does not appear in the multi-dimensional space, but is an artifact of the 2D embedding. However, the problem of minimizing the number of edge crossings is NP-hard~\cite{garey1983crossing}. Therefore, we opt for a suitable approximation algorithm using a planarization method~\cite{tamassia2013handbook}. For simplicity of the rasterization in the next steps, we use an orthogonal graph drawing algorithm that assures that the boundary vertex $v_b$ is placed close to the boundary. We then create a rasterization of the graph embedding to create an initial configuration for the cellular automaton. An example of such an initial configuration is shown as the graph embedding in \cref{fig:overview}. More details on the graph embedding and creating the initial configuration can be found in the supplementary material.
While the graph drawing already provides a 2D embedding of the segmentation's topology, it does not yet reflect the segment sizes and boundary sizes, which we will add next.

\subsection{Cellular Automaton}
\label{sec:automaton}
The graph embedding places the graph vertices in a 2D space and draws connecting lines for all graph edges. We aim for a visualization that resembles a 2D segmentation visualization such that the mental abstraction to other representations, such as graph drawings, is not necessary.  Using the graph embedding as an initial configuration, we want to transform it to a representation where (1) the graph vertices $v_i \in V$ become 2D segments with an area according to the vertex weights $A_i$ and (2) the graph edges $(v_i,v_j)\in E$ become shared segment boundaries between the 2D segments belonging to $v_i$ and $v_j$ with a boundary length according to the edge weight $B_{ij}$ (cf.~\cref{fig:overview}). The first goal is related to that of generating cartograms. We, therefore, build upon an approach by Dorling~\cite{dorling1996area} using a cellular automaton. While this approach was criticized for not preserving shapes~\cite{nusrat2016state}, this is not an issue for us, as we cannot, generally, preserve the shape of $n$-dimensional segments in 2D. Moreover, we extend the approach by Dorling to also fulfill the second goal of boundary length optimization listed above.

The cellular automaton consists of cells arranged in a regular 2D grid. Each cell of the cellular automaton has a state that can change based on a set of rules that depends on the state of the neighboring cells. An iterative process allows for the optimization of a target function. For our application, each cell's state corresponds to one of the segment IDs. If the cell is not assigned to any segment (in the following referred to as background cells), we assign the ID $-1$. If the cell represents an edge crossing (black dots in \cref{fig:overview}), we assign the ID $-2$. 
Given the initial configuration, the cellular automaton iteratively changes the state of the cells facilitating the local optimization according to some optimization criteria.

\smallskip
\noindent\textbf{Optimization Criteria.}
When applying the cellular automaton approach to the initial configuration, each cell can change its state (i.e., its label ID) based on the states of the surrounding cells in each iteration. 
During the iterative process, we want to optimize the area of the segments and the shared boundary lengths between segments while preserving the topology. 

To optimize the \textit{segments' areas}, we compute for each cell the deviation of the size of the 2D segment it belongs to from the size of the $n$-dimensional segment that it corresponds to. 
The deviation for segment $s$ is computed by
\begin{equation}
\label{eq:areaDeviation}
  d_{A,s}=\frac{A_{s,nD}}{A_{nD}} - \frac{A_{s,2D}}{A_{2D}}\ ,
\end{equation}
where $A_{s,nD}$ denotes the size of the $n$-dimensional segment that corresponds to $s$, $A_{nD}$ the overall size of the partitioned multi-dimensional domain, $A_{s,2D}$ the current size of segment $s$ in the 2D embedding (computed as the number of cells assigned to segment $s$), and $A_{2D}$ the size of the 2D embedding (computed as the total number of cells of the cellular automaton). Here, $d_{A,s}>0$ denotes that the segment $s$ needs to expand further to represent the original partitioning accurately. 
Having computed the deviation $d_{A,s}$ for all segments $s$, we determine whether a cell should change its state by investigating, if at least one of the neighboring cells has a higher deviation. A cell with a neighbor with higher area deviation will be marked for a state change from its current segment ID to the neighboring segment ID.

\begin{figure}[tb]
     \centering
     \begin{subfigure}[b]{0.31\textwidth}
         \centering
         \includegraphics[width=\textwidth, alt={Two subfigures with 9 pixels each. In the first ones, three boundaries have different colors and are marked with a gray cross. In the second, four boundaries have different colors and are marked with gray crosses.}]{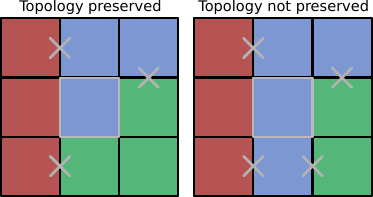}
         \caption{Number of changes}
         \label{fig:topologyPreservation}
     \end{subfigure}
     \hfill
     \begin{subfigure}[b]{0.16\textwidth}
         \centering
         \includegraphics[width=\textwidth, alt={Every second pixel in every second row is black, the others are light gray.}]{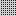}
         \caption{Changing Pattern}
         \label{fig:ourCheckerboard}
     \end{subfigure}
        \caption{(a) Topology is preserved, if the number of segment changes (grey crosses) when clockwise traversing the Moore neighborhood of the current cell (grey boundary) is $\leq3$. Otherwise, topology might change. (b) To prevent topology violations by simultaneously changing neighboring cells, only non-adjacent cells may change simultaneously.}
\end{figure}

To optimize for shared \textit{boundary lengths}, we employ a similar approach. However, the boundary deviation is not computed per segment, but per pair of segments. In analogy to the segment size deviation, we compute the boundary length deviation as the difference of the relative boundary length in high-dimensional space to the relative boundary length in the 2D embedding. Thus, the boundary length deviation between segments $s_i$ and $s_j$ is computed by 
\begin{equation}
\label{eq:boundaryDeviation}
  d_{L,(s_i,s_j)}=\frac{L_{(s_i,s_j),nD}}{L_{nD}} - \frac{L_{(s_i,s_j),2D}}{L_{2D}}\ ,
\end{equation}
where $L_{(s_i,s_j),nD}$ denotes the boundary length between the $n$-dimensional segments corresponding to segments $s_i$ and $s_j$, $L_{nD}$ the sum of the lengths of all boundaries in the $n$-dimensional space, $L_{(s_i,s_j),2D}$ the boundary length between segments $s_i$ and $s_j$ in the 2D embedding, and $L_{2D}$ the sum of the lengths of all boundaries in the 2D embedding. Here, $d_{L,(s_i,s_j)}>0$ denotes that the length of the boundary needs to increase. Note that for each cell, we can compute a boundary deviation for each neighbor leading to four boundary deviations per cell. However, just changing the segment to which the cell belongs does not necessarily change the length of the boundary. For example, the boundary length between the blue and the green segment in \cref{fig:topologyPreservation} is identical even though one cell has a different state. 
Therefore, we compute the change in boundary length $L_{(s_i,s_j),2D}$ when changing the label of the current cell from segment $s_i$ to segment $s_j$ by
$
    \Delta L_{(s_i,s_j)} = N_{s_i}-N_{s_j},
$
where $N_{s_i}$ is the number of neighbors of the current cell that belong to segment $s_i$ and $N_{s_j}$ the number of neighbors of the current cell that belong to segment $s_j$.
Then, the cell is marked for a state change, if and only if $d_{L,(s_i,s_j)}\Delta L_{(s_i,s_j)}>0$. In this case, the boundary of the 2D segment is proportionally larger than in the multi-dimensional space  ($d_{L,(s_i,s_j)}<0$) and the boundary length would be reduced ($\Delta L_{(s_i,s_j)}<0$), or the boundary is too short ($d_{L,(s_i,s_j)}>0$) while the length of the boundary can be expanded ($\Delta L_{(s_i,s_j)}>0$).

If a cell was marked for a state change, the \textit{target segment} to which the cell should change needs to be defined. Here, we check the area as well as the boundaries to the four neighbors of the cell and choose the criterion with the most significant deviation to determine the target segment of the cell. Note that the initial configuration contains background cells that we want to vanish. Thus, we set their area deviation to $-1$ independent of the background's size. This corresponds to the minimum value and would push the cellular automaton towards changing the background cells' state to other segments, leading to a vanishing background.

While these optimization criteria do not consider the shape of the segments, more compact shapes are, generally, easier to interpret than line-like structures. To achieve compact shapes, we use the \textit{security factor} introduced by Dorling~\cite{dorling1996area}. The security factor considers the cell's neighborhood and measures how exposed the cell is to the corresponding segments. Each von Neumann neighbor that belongs to the same segment adds a value of $3$ to the security factor. Here, we also consider the neighbors on the diagonal, which add a value of $1$. Thus, we obtain a value between $0$ for an isolated cell label and $16$ for a cell completely surrounded by cells of the same segment. We only allow cells to change their state, if the security factor lies below the threshold of $11$. A detailed investigation of the choice of the security factor and its influence on the result can be found in \cref{sec:parameters}.

\smallskip
\noindent\textbf{Topology Preservation.}
Besides optimizing the areas and boundary sizes, the cellular automaton should preserve the topology, for which we follow the approach by Dorling~\cite{dorling1996area} with some adaptations. It can be tested if a cell is critical to preserve the topology by counting the number of segment changes that occur when traversing the cell's Moore neighborhood, i.e., its 8-connected neighbors, in clockwise order. If the number of segment changes is larger than $3$, the cell is critical for the topology and should not be changed, see \cref{fig:topologyPreservation}. To preserve the topology on the grid's border, we also count the transition to the border as a segment change.
Additionally, isolated cells should not be allowed to disappear. As we initially start with a background that should disappear over time, we need to adapt the behavior for the background cells. Note that only changing the state of background cells cannot change the topology because of the topology preservation criterion: If the state change of a background cell leads to a new shared boundary of two segments, the number of segment changes when traversing the background cell's neighborhood must be at least $4$ and is therefore marked as critical. This restriction could lead to background cells not vanishing during the iterative process. Therefore, we separately check for the background cells to investigate, if a change would destroy the topology based on the original graph. In some cases, especially for complex, non-planar graphs, some of the background cells might remain to separate two segments. As removing them would lead to topological changes, we keep them as separating boundaries and treat them differently in the visualization to distinguish them from the segments (see \cref{sec:visualization}).
Note that the cells that denote an edge crossing are never considered for change because, by definition, they contain at least $4$ changes when traversing the cell's neighborhood.

\smallskip
\noindent\textbf{Iterative Optimization.}
Following the criteria described above, the 2D embedding is generated iteratively. To avoid conflicts by simultaneously changing neighboring cells, Dorling~\cite{dorling1996area} proposed a \textit{checkerboard pattern} where diagonal neighbors change simultaneously. However, applying a standard checkerboard pattern might lead to topology violations in the Moore neighborhood. Even though we compute the topology for the von Neumann neighborhood, we want to avoid diagonal neighbors in the embedding to avoid confusions. Thus, we apply a changing pattern that does not allow direct or diagonal neighbors to change simultaneously, which leads to a pattern as shown in \cref{fig:ourCheckerboard}. This pattern is iteratively shifted such that each cell may be changed in every fourth iteration.

Using an iterative algorithm, we need to define some \textit{convergence criterion} to determine when the algorithm should terminate. In general, the user can set a maximum number of iterations. However, we want the cellular automaton to terminate early, if no further improvements can be made. For this purpose, we check after each iteration, if the states changed. If no further changes occur for four iterations, the algorithm has converged. However, one may observe an oscillation between different states without improvement. Therefore, we apply a damping that gets stronger with increased accuracy. 
For the damping, we define a probability that a state switches. This probability is defined as the absolute value of the maximal deviation of either the areas or the boundaries multiplied by a user-defined scaling factor $g$. A higher value of $g$ indicates a higher tolerance to the deviation and, thus, a weaker damping. Then, the cell switches its state with the calculated probability. However, the stochastic nature of this approach might cause the algorithm to stop early. To prevent this, we change the threshold for stopping the algorithm to ten iterations without changes, which is also a parameter that can be adjusted by the user. 

\begin{figure}[tb]
     \centering
     \begin{subfigure}[b]{0.24\textwidth}
         \centering
         \includegraphics[width=\textwidth, alt={Segmentation embedding with three edge crossings marked by black circles and one edge crossing marked by a gray circle.}]{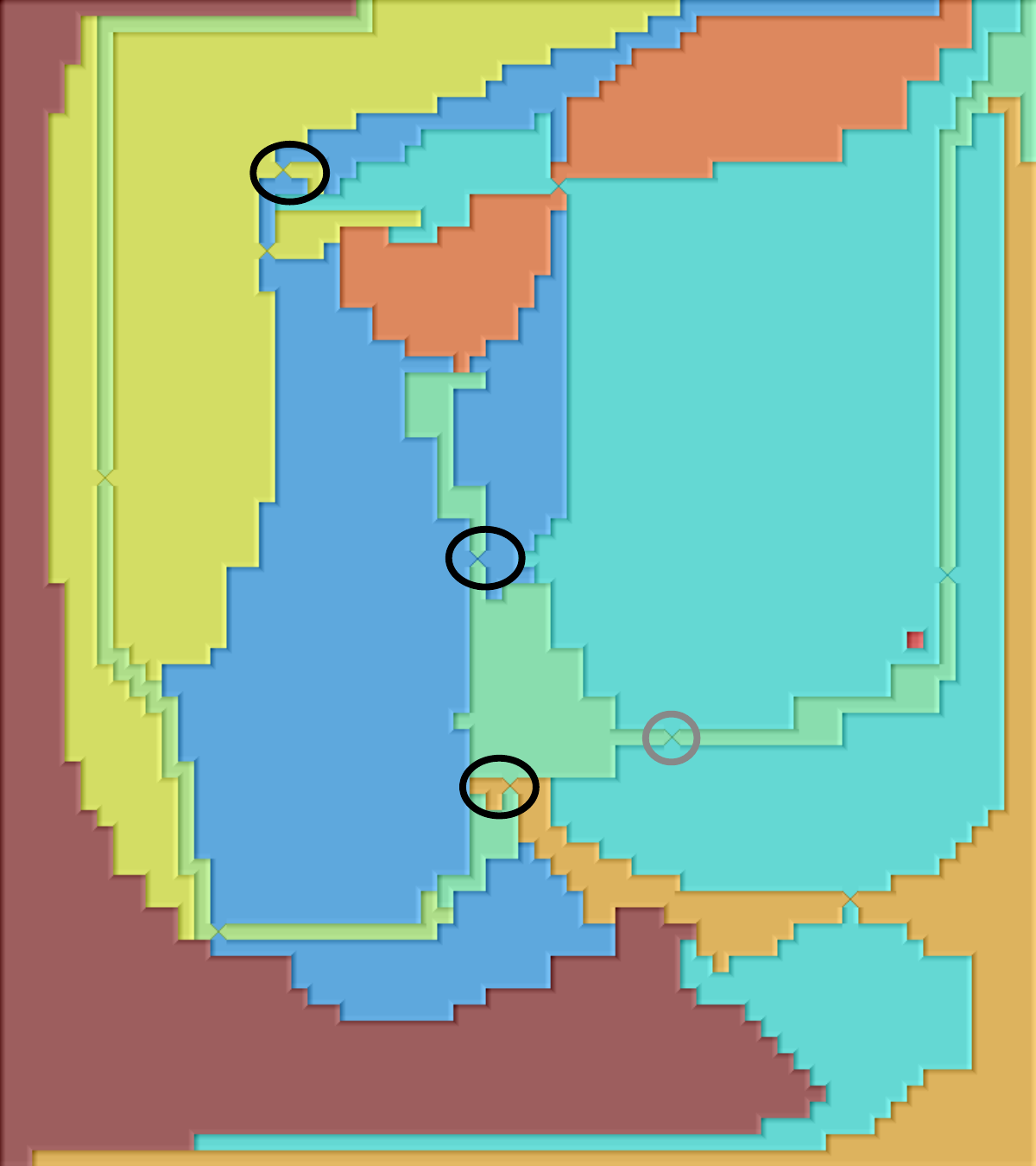}
         \caption{Before}
         \label{fig:removalBefore}
     \end{subfigure}
     \hfill
     \begin{subfigure}[b]{0.24\textwidth}
         \centering
         \includegraphics[width=\textwidth, alt={Segmentation embedding with three positions of former edge crossings marked by black circles and one position of former edge crossing marked by a gray circle. These three edge crossings have been removed and the respective segments are replaced by white background pixels.}]{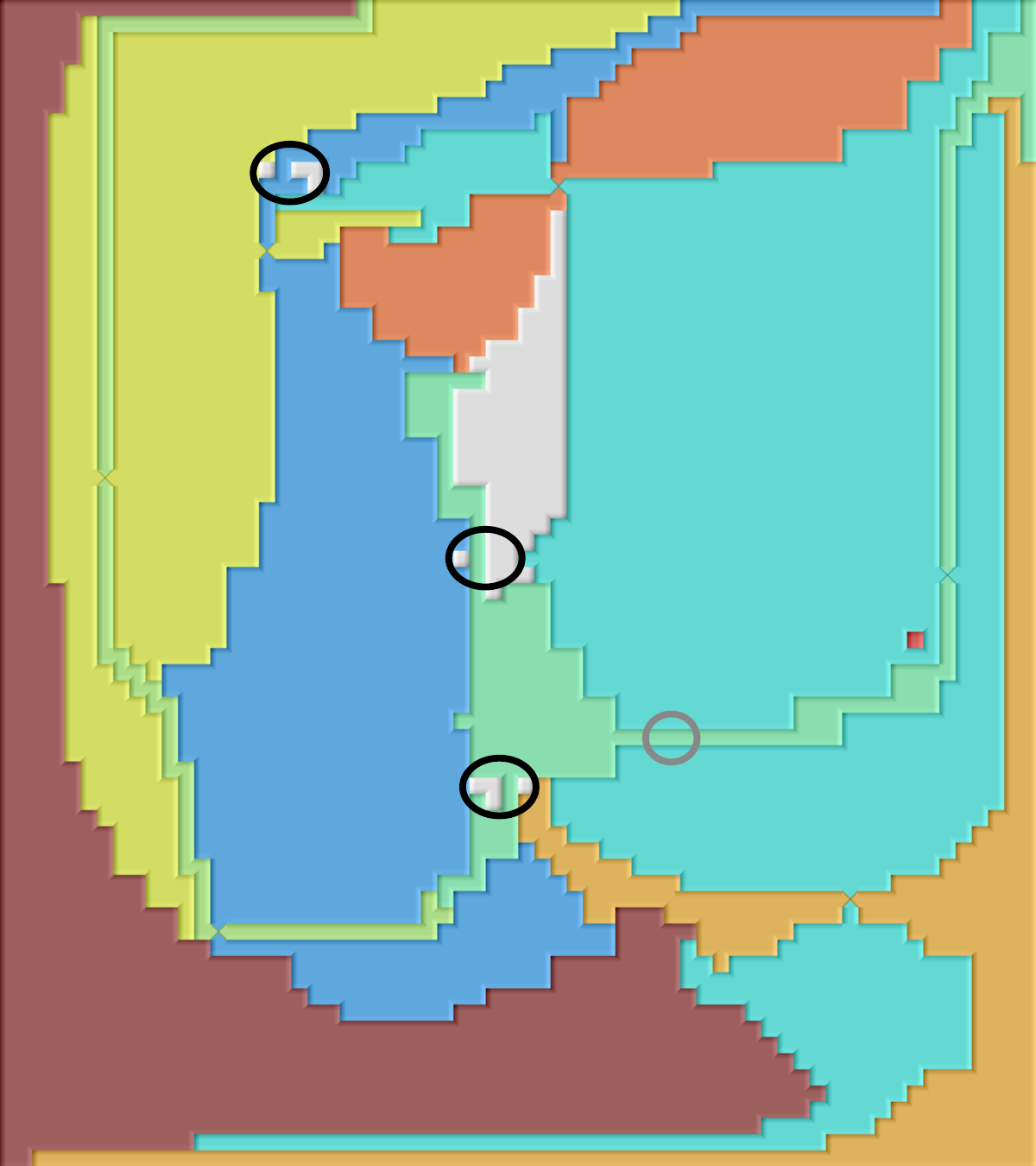}
         \caption{After}
         \label{fig:removalAfter}
     \end{subfigure}
        \caption{To improve the embedding, crossings between segments that become unnecessary can be removed. These crossings are either duplicates (grey circle) or crossings where one side is not required anymore to preserve the topology (black circles).}
        \label{fig:removal}
\end{figure}

As we are using a graph layout for node-link diagrams but then generate a dense visualization of the segmentation embedding instead, it may occur that some \textit{edge crossings} created during the initial graph layout are no longer necessary. Here, we can differentiate between two cases: First, two segments could cross twice, where one of the crossings could be removed (see grey circles in \cref{fig:removal}), but we need to ensure that none of the segments is split. Second, an edge crossing was necessary to connect to another segment or the border, but due to the disappearance of the background, this connection is now occurring in another part of the segment. In this case, the part of the segment whose topological information can be fully replicated by the other part can be removed together with the corresponding edge crossing (see black circles in \cref{fig:removal}). This is implemented by setting the states of the corresponding cells back to the background, which creates additional space for the surrounding segments to spread further. As testing the conditions for removal is computationally expensive, we do not perform this step every iteration but in a user-defined frequency. For our results, we compute the check every $300$th iteration.

\begin{figure}
     \centering
     \begin{subfigure}[b]{0.24\textwidth}
         \centering
         \includegraphics[width=\textwidth, alt={A segmentation embedding with a complex nested structure.}]{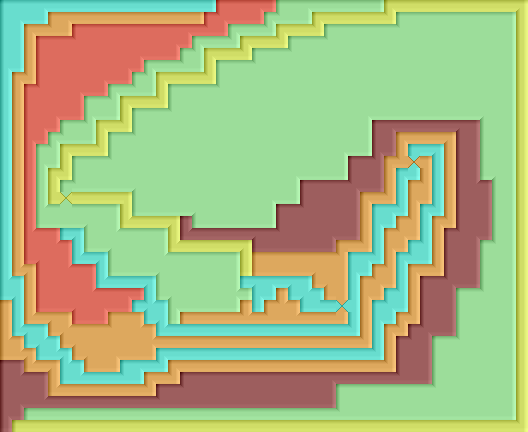}
         \caption{Fixed crossing positions}
         \label{fig:withoutMoving}
     \end{subfigure}
     \hfill
     \begin{subfigure}[b]{0.24\textwidth}
         \centering
         \includegraphics[width=\textwidth, alt={A segmentation embedding with a less nested structure.}]{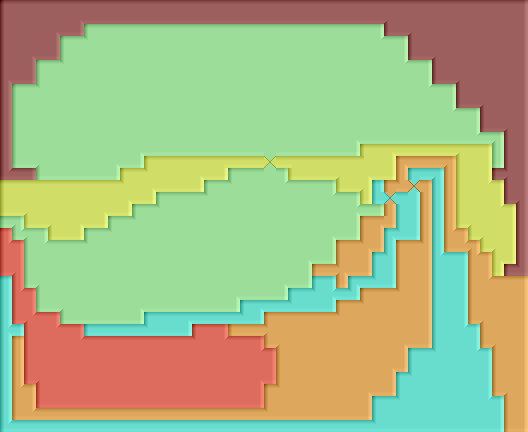}
         \caption{Flexible crossing positions}
         \label{fig:withMoving}
     \end{subfigure}
    \caption{Keeping the edge crossing position fixed might result in unnecessarily complex embeddings (a), while allowing them to move leads to more compact segments (b). Both examples have been created based on the same initialization.}
    \label{fig:moving}
\end{figure}

The \textit{positions of the edge crossings} are fixed in the layout. However, allowing them to change their positions can significantly improve the layout as shown in \cref{fig:moving}. Keeping the line crossings fix provides less flexibility for the relaxation and, thus, might induce more line-like structures. Thus, we allow them to move by switching the state of the cell containing the edge crossing with the state of a neighboring cell. The movement is only allowed, if the topology is preserved. If a vertical or horizontal movement is possible, we compute the barycenter of the involved segment and move the edge crossing in the direction towards the segment's barycenter, as moving towards the barycenter generally allows for more compact segments.

The evolution of our algorithm over time, including the different additional improvements discussed in this section, is shown in an animation in the supplementary material.

\section{Rendering}
\label{sec:visualization}
\begin{figure}
    \centering
    \includegraphics[width=0.8\linewidth, alt={A function over a line of pixels. At the boundary between pixels, the function decreases to zero. At the segment indicating background pixels, the function decreases to -1.}]{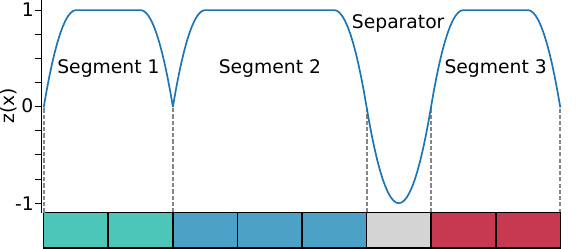}
    \caption{We create a height profile with plateaus in the center of the segments and valleys for separators between segments. Here, we set the height $h=1$ and the width for height changes $w=0.5r$, where $r$ is the number of pixels for each cell of the cellular automaton.}
    \label{fig:shading}
\end{figure}
Inspired by the renderings of 2D segmentations, we visualize the 2D embedding by assigning to each segment one \textit{color}. The colors may be assigned according to some categorical color map (i.e., trying to use distinguishable colors) but may also encode some other (meta-)information about the data, for example, indicating which segments lead to a similar simulation output when analyzing multi-dimensional parameter space partitionings. To still clearly observe the segments even in case of adjacent segments with similar colors, we emphasize boundaries by applying a \textit{shading} inspired by cushion treemaps~\cite{van1999cushion, telea2014data}. The shading should allow us to identify the unique segments independent of the color coding to provide a large flexibility for using different colors. Additionally, the shading should allow for a better differentiation of the separators  (regions of background cells that remain to separate different segments) between segments and shared segment boundaries. 
Unlike cushion treemaps, we do not want to visualize a hierarchy, but aim at a relatively simple visualization without adding too much visual complexity. 
We follow the idea of cushion treemaps and use quadratic functions to model the height, but use a piecewise definition to keep the inner part of the segments flat. 

Our piecewise height function $z_i(x)$ in one dimension is shown in \cref{fig:shading}. We use an increase in height from the shared boundaries of the different segments, while we use a decrease for the separators. We also choose a color for the separators that is clearly distinguishable from all segment colors. 

In the following, we will describe our shading based on the $x$-direction of the image. The $y$-direction is treated identically. To model the desired behavior, we apply the constraints $z_i(x_{i,1})=0$, $z_i(x_{i,1}+w)=h$ and $\frac{\mathrm{d}z_i}{\mathrm{d}x}(x_{i,1}+w)=0$ where $w$ is the width of the region where the quadratic increase in height should occur, $x_{i,1}$ is the position where the segment $i$ starts, and $h$ denotes the final height of the segments. Analogous constraints are applied at the end of the segment, where the height decreases. For the separators, which are modeled as valleys with a negative height, we set the constraints $z_v(x_{v,2}-w) = -h$, $\frac{\mathrm{d}z_v}{\mathrm{d}x}(x_{v,2})=0$ and $\frac{\mathrm{d}z_v}{\mathrm{d}x}(x_{v,2})=\frac{\mathrm{d}z_i}{\mathrm{d}x}(x_{i,1})$ where $x_{v,2}$ denotes the position where the separator $v$ ends. Again, analogous constraints are applied at the start of the separators. These constraints can be used to compute the normals in these regions as $(2(ax+b), 2(ay+c), 1)$, where $a$, $b$, and $c$ are the coefficients that  can be computed between the boundaries of the segments or separators $x_1$ and $x_2$ in $x$-direction and $y_1$ and $y_2$ in the y-direction as follows:
\begin{equation*}
  a=\begin{cases}
      0 & x_1+w < x < x_2-w \\
      -\frac{sh}{w^2} & \mathrm{otherwise}
      \end{cases}\ ,
\end{equation*}
\vspace{-0.2cm}
\begin{equation*} 
  b=\begin{cases}
      \frac{sh}{w^2}(x_1+w) & x_1 \leq x \leq x_1+w \\
      0 & x_1+w < x < x_2-w \\
      \frac{sh}{w^2}(x_2-w) & x_2-w \leq x \leq x_2
      \end{cases}\ ,
\end{equation*}
\vspace{-0.1cm}
\begin{equation*} 
  c=\begin{cases}
      \frac{sh}{w^2}(y_1+w) & y_1 \leq y \leq y_1+w \\
      0 & y_1+w < y < y_2-w \\
      \frac{sh}{w^2}(y_2-w) & y_2-w \leq y \leq y_2
    \end{cases}\ ,
\end{equation*}
where $s=1$, if the corresponding region is a segment, and $s=-1$, if the region corresponds to a segment separator.

\begin{figure}[tb]
    \centering    \includegraphics[width=\linewidth, alt={Big segmentation embedding where each segment is colored in red. The boundaries are shown by a perceived decrease in the surface. An inset shows a zoom-in view of an edge crossing and a separator.}]{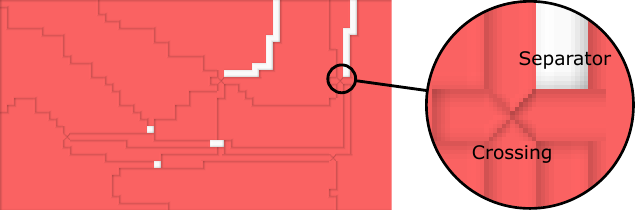}
    \caption{Shading the segments and visually encoding special features like edge crossings and background areas allows for a visualization of segments independent of the color coding.}
    \label{fig:rendering}
\end{figure}

By applying this kind of rendering, a wide range of color coding options can be used. \cref{fig:rendering} shows a visualization where all segments contain the same color. Still, the different segments and the edge crossings are visible.
The \textit{edge crossings} are visually encoded as a cross where we apply a similar shading as for the segment boundaries. Thus, the crossings are visible but are not highlighted.

\section{Algorithmic Evaluation}
\label{sec:evaluation}
In the following, we will investigate the influences of the different optimization criteria of our approach as well as its scalability with the number of segments and the dimensionality of the input data. Two synthetic datasets allow us to tune the characteristics as desired.

As a first dataset $D_1$, we create a 2D or 3D image segmentation with a user-defined resolution, an adaptable number of segments, and varying segment shapes.  
For each segment, we randomly choose a seed point and a random growth rate. Then, each segment is expanded by adding surrounding, still unassigned cells using the growth rate as a probability. 
As a second dataset $D_2$, we use a simple 3D cube that is divided once in all dimensions leading to eight, equally sized segments where each segment has a joint boundary with three other segments and the outer boundary of the domain, see \cref{fig:overview}.

\label{sec:qualityCriteria}
To evaluate our approach with respect to different influencing factors, we define a set of quality metrics that we want to investigate: \\
\noindent\textit{Number of edge crossings:} We investigate the absolute number of edge crossings, as fewer edge crossings mean a less complex embedding. \\
\noindent\textit{Mean area deviation:} We measure how well our area optimization works by computing the mean area deviation for all segments $\Bar{d_A}=\frac{1}{|S|}\sum_{s \in S}|d_{A,s}|$, where $d_{A,s}$ is computed as in \cref{eq:areaDeviation} and $|S|$ denotes the cardinality of $S$. \\ 
\noindent\textit{Mean boundary length deviation:} We measure how well our boundary length optimization works by computing the mean boundary length deviation for all adjacent segments $s_i$ and $s_j$ by $\Bar{d_L}=\frac{1}{|E|}\sum_{(s_i,s_j) \in S\times S}|d_{L,(s_i,s_j)}|$, where $d_{L,(s_i,s_j)}$ is the boundary length deviation as in \cref{eq:boundaryDeviation} and |E| denotes the number of edges in the graph representing the segmentation.

\noindent
Note that the topology preservation is not explicitely evaluated because the topology is perfectly preserved based on the construction of the algorithm.

\begin{figure}[tb]
     \centering
     \begin{subfigure}[b]{0.22\textwidth}
         \centering
         \includegraphics[width=\textwidth, alt={2D segmentation where each segment has its own color..}]{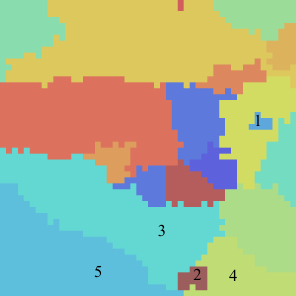}
         \caption{Original}
         \label{fig:2dOriginal}
     \end{subfigure}
     \hfill
     \begin{subfigure}[b]{0.25\textwidth}
         \centering         \includegraphics[width=\textwidth, alt={2D segmentation with the rendering for segmentation embeddings. The same colors as in (a) are used.}]{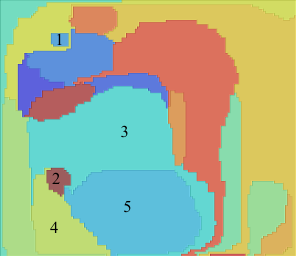}
         \caption{Embedded}
         \label{fig:2dEmbedded}
     \end{subfigure}
    \caption{A 2D embedding (b) of a 2D segmentation (a) allows for a visual comparison of input and output of our algorithm. We observe that the topology is maintained (including the light blue segment (label 1) fully surrounded by the yellow one) and that area sizes and shared boundary lengths are approximately the same. Shapes and locations are not preserved, but this was not a desired goal of our approach.}
    \label{fig:2dExample}
\end{figure}

\subsection{Visual Inspection and Numerical Evaluations}
We first apply our algorithm to the 2D version of dataset $D_1$ with 20 segments. Using a 2D input is obviously not a meaningful application scenario, but it allows for a visual inspection and comparison of input and output, which are shown in \cref{fig:2dExample}. The output is obtained after $5,000$ iterations with a damping factor of $7$ and a security factor of $11$. We observe that topological structures are preserved. Moreover, area sizes in both images are similar, which is confirmed by a mean area deviation of $\Bar{d_A} = 0.016\%$. Also, boundary lengths are similar, which is confirmed by a mean boundary length deviation of $\Bar{d_L}=0.96\%$. For example, when investigating the brown segment in the lower left part of the embedding (label 2), most of its boundary is shared with the lighter blue (label 3) and the green segment (label 4), while only a small shared boundary appears with the darker blue color (label 5), which agrees well with the original segmentation. Location of segments and shapes are not preserved, but this was not a desired goal of our approach.

Next, we apply our approach to dataset $D_2$ using the same parameters as above, see \cref{fig:overview}. Here, we obtain a mean area deviation of $\Bar{d_A}=0.18\%$ and a mean boundary length deviation of $\Bar{d_L}=4.49\%$. In general, the deviations are significantly higher than for the 2D case. However, this was to be expected, as in this case, a dimensionality reduction is involved and, additionally, the graph cannot be embedded in 2D as a planar graph. Therefore, we obtain four edge crossings and some separating regions of segments without a shared boundary. 

The quality measures for all datasets discussed in this paper are presented in Table 1 of the supplementary material. Here, we can observe that the results are rather dataset-dependent.

\subsection{Influence of Parameters}
\label{sec:parameters}
\begin{figure*}[htb]
    \centering
    \includegraphics[width=\linewidth, alt={Boxplots showing the mean area dedication and the mean boundary deviation for different damping factors and security factors. Most boxplots cover a similar range of data and have some outliers.}]{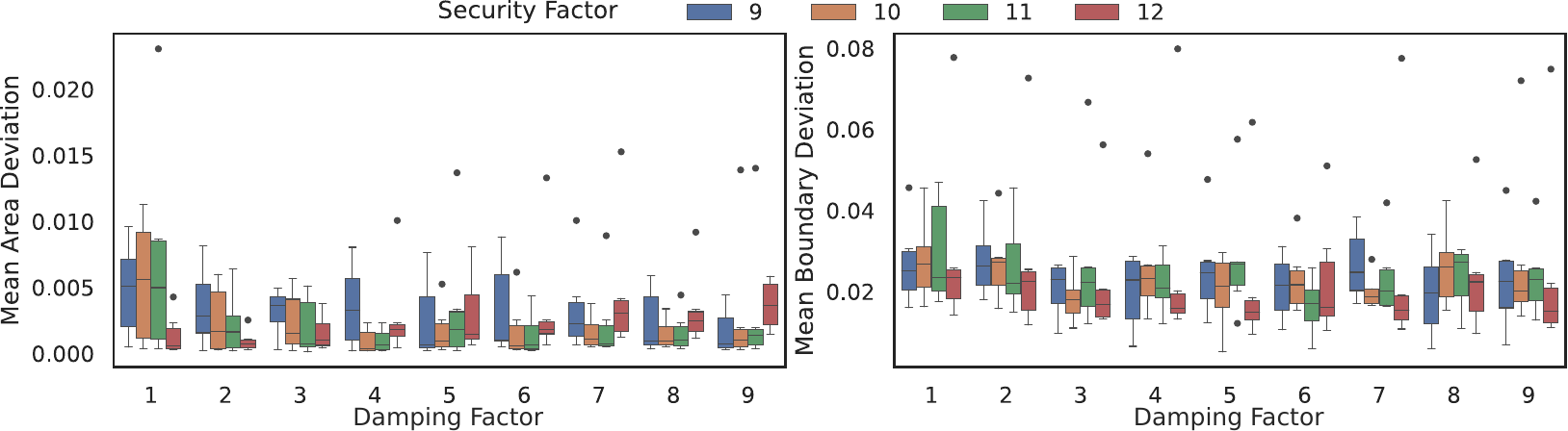}
    \caption{Mean area deviation and mean boundary length deviation are evaluated depending on the damping factor as well as the security factor.}
    \label{fig:parameters}
\end{figure*}
Two main parameters for our algorithm are the choice of the damping factor and the security factor. To identify their influence on the outcome, we study damping factors between $1$ and $9$ and security factors between $9$ and $12$. We use dataset $D_1$ with $5$, $10$, and $15$ segments in 2D and 3D settings. For all cases, we set the number of iterations to the number of pixels as the image resolution scales with the segmentation's complexity to visualize.
The results are shown in \cref{fig:parameters}. We can observe that the accuracy slightly increases for damping factors $>3$. This observation can be explained by the fact that too strong damping leads to early termination of the algorithm, i.e., the result is not fully optimized. Therefore, we recommend a larger damping factor, even though it depends on the underlying dataset. When observing the results for the different security factors, we can observe that $10$ and $11$ lead to better results in mean area deviation. The variation in mean boundary length deviation is less pronounced. Here, the security factor of $12$ outperforms the others, which is expected as it allows for more flexibility in forming boundaries. However, as it performs worse in area deviation, we still recommend security factors of $ 10 $ or $ 11 $ based on the numerical results.
\begin{figure*}[htb]
     \centering
     \begin{subfigure}[b]{0.24\textwidth}
         \centering
         \includegraphics[width=\textwidth, alt={Segmentation embedding with relatively straight boundaries.}]{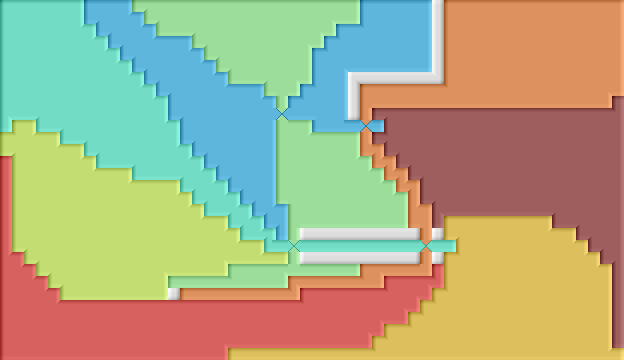}
         \caption{Security factor 9}
         \label{fig:secFactor9}
     \end{subfigure}
     \hfill
     \begin{subfigure}[b]{0.24\textwidth}
         \centering
         \includegraphics[width=\textwidth, alt={Segmentation embedding with relatively straight boundaries.}]{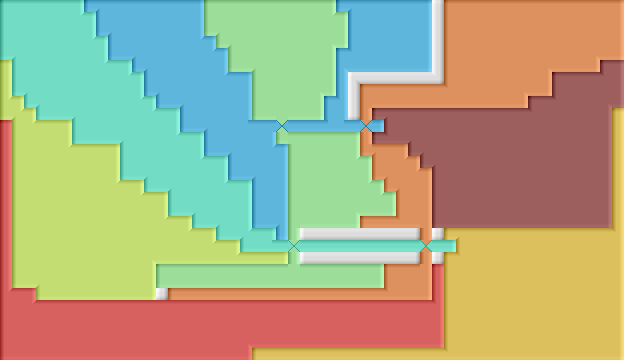}
         \caption{Security factor 10}
         \label{fig:secFactor10}
     \end{subfigure}
     \hfill
     \begin{subfigure}[b]{0.24\textwidth}
         \centering
         \includegraphics[width=\textwidth, alt={Segmentation embedding with relatively straight boundaries.}]{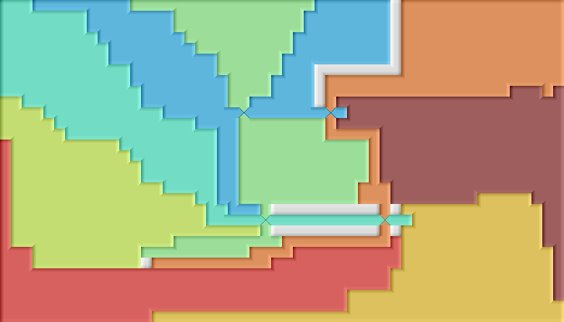}
         \caption{Security factor 11}
         \label{fig:secFactor11}
     \end{subfigure}
     \hfill
     \begin{subfigure}[b]{0.24\textwidth}
         \centering
         \includegraphics[width=\textwidth, alt={Segmentation embedding with complex boundaries.}]{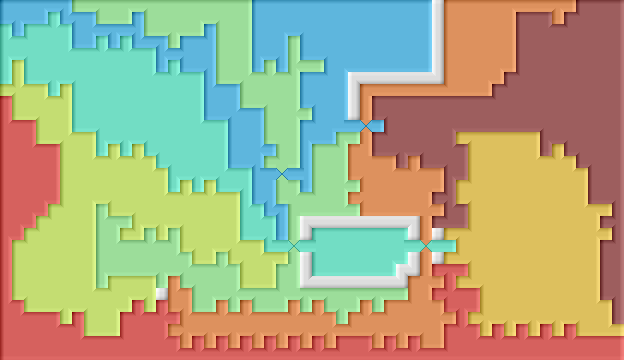}
         \caption{Security factor 12}
         \label{fig:secFactor12}
     \end{subfigure}
        \caption{Cellular automaton with different security factors applied to the same initial state: The result for a security factor of $12$ differs significantly from the results with lower values, introducing unnecessary visual complexity.}
        \label{fig:securityFactors}
\end{figure*}

The visual differences between the security factors are shown in \cref{fig:securityFactors}. The results for security factors $9$ to $11$ are quite similar and the visual differences are negligible. However, for a security factor of $12$, we see that the boundaries between the segments are significantly more fine-grained. As this adds additional complexity to the visualization and, as seen in \cref{fig:parameters}, does not improve the numerical results in general, we recommend choosing a smaller security factor. For the following results, we use a security factor of $11$ as a default.

\subsection{Optimization Criteria}
\begin{figure}[htb]
    \centering
    \includegraphics[width=\linewidth, alt={First part: two boxplots for the mean area deviation with a p-value of the paired t-test of 0.048. Second part: two boxplots for the mean boundary deviation with a p-value of the paired t-test of 0.14. Third part: Two boxplots for the time with a p-value of the paired t-test of 0.016.}]{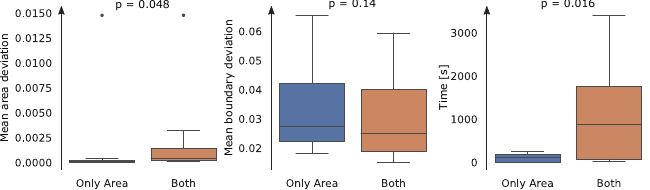}
    \caption{The comparison of different evaluation criteria shows the tradeoffs between only optimizing for area and optimizing for both area and boundary length. The p-values are computed using a paired t-test.}
    \label{fig:optimization}
\end{figure}
Next, we want to evaluate the impact of area and boundary length preservation. Only optimizing boundary length does not lead to desirable results, as the boundary length optimization does not allow for a spreading towards the background voxels of the initial graph embedding.
Hence, we compare using area optimization only to the combined optimization of area and boundary length with respect to the quality metrics defined above.
We choose dataset $D_1$ with $5$ to $15$ segments in a 3D setting and a security factor of $11$.  Note that the area deviations are significantly smaller than the boundary length deviations leading to a smaller mean error when considering only areas. To avoid early stopping in this case, we increase the damping factor to $100$ in this evaluation (while choosing $7$ in the other cases). The results are shown in \cref{fig:optimization}. We observe that adding boundary length optimization leads to a larger area deviation, but to a significant decrease in boundary deviation as expected. However, the difference for the dataset variants investigated here is relatively small. At the same time, the computations take much longer if the boundary deviation is included. Thus, for larger graphs with more complex embeddings when slightly larger boundary deviations may be tolerable, we recommend only optimizing for areas.

\subsection{Scalability}
The computational scalability of our approach depends on different factors. The first step is the graph computation. Computing the vertices including their weights scales linearly with the number of cells $N$ in the multi-dimensional domain, leading to a complexity of $\mathcal{O}(N)$. For determining shared boundaries, we obtain a complexity of $\mathcal{O}(Nn)$ where $n$ is the number of dimensions, because, for each voxel, both neighbors in each dimension need to be checked. Note that the number of cells increases exponentially with the number of dimensions (when assuming the same sampling). The subsequent steps for creating the layout only depend on the weighted graph. Thus, the remainder of the algorithm is insensitive to the resolution and dimensionality of the input data, but instead scales with its internal complexity, which is characterized by the number of vertices and edges. A more complex graph is not only more expensive to draw, but also requires a larger number of cells for the cellular automaton, which leads to larger computation times in the cellular automaton. Additionally, edge crossings increase the computational cost as additional optimization steps are executed. 
This can be confirmed experimentally. We observe that the computational costs increase significantly with the number of segments (see \cref{fig:scaling} in the supplementary material). The increasing number of cells most likely also causes the decrease in the mean deviation.

Table 2 of the supplementary material provides an overview of the run times for the different steps of the algorithm of different real-world and artificial datasets. Which step is the computationally most expensive one strongly depends on the dataset. The computation time for the graph is mainly determined by the number of voxels in the input space, which confirms the theoretical considerations. The complexity of the graph, for which the number of edge crossings and the number of segments provide an indicator, mainly determines the computation time for the embedding. The number of cells of the cellular automaton is the main influencing factor for the computational cost of running the automaton. However, early termination of the automaton significantly reduces the computation times for this step. The respective quantitative analysis is detailed in Section~\ref{sec:appendixEvaluation} of the supplementary material.

\section{Application to Real-world Scenarios}
\label{sec:application}
We demonstrate the applicability of our approach to real-world problems by applying it to the visualization of 3D image segmentations as well as to 4D and 5D parameter space partitionings of simulation ensembles. We also discuss our approach in comparison to visualizing the segmentation as a graph drawing. 
A second 3D segmentation use case can be found in the supplementary material.

\begin{figure*}[hbt]
     \centering
     \begin{subfigure}[b]{0.2\textwidth}
         \centering
         \includegraphics[width=\textwidth, alt={Segmentation rendering for the needle, the tumor and the vessels in the 3D domain.}]{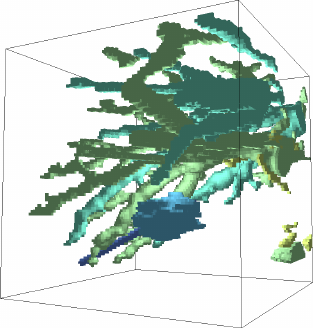}
         \caption{Volume visualization}
         \label{fig:ablationVolume}
     \end{subfigure}
     \begin{subfigure}[b]{0.39\textwidth}
         \centering
         \includegraphics[width=\textwidth, alt={Segmentation embedding where each segment is colored according to the tissue type. The individual segments are labeled by the tissue type to provide a color map.}]{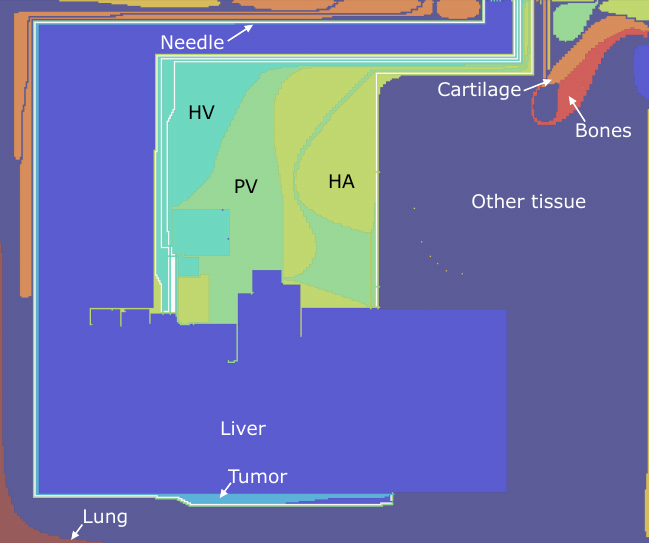}
         \caption{Segmentation embedding}
         \label{fig:ablationEmbedding}
     \end{subfigure}
          \begin{subfigure}[b]{0.33\textwidth}
         \centering
         \includegraphics[width=0.98\textwidth, alt={Node-link diagram of the segmentation embedding using the same colors as b. Red boxed highlight nodes that are not visible due to their small size.}]{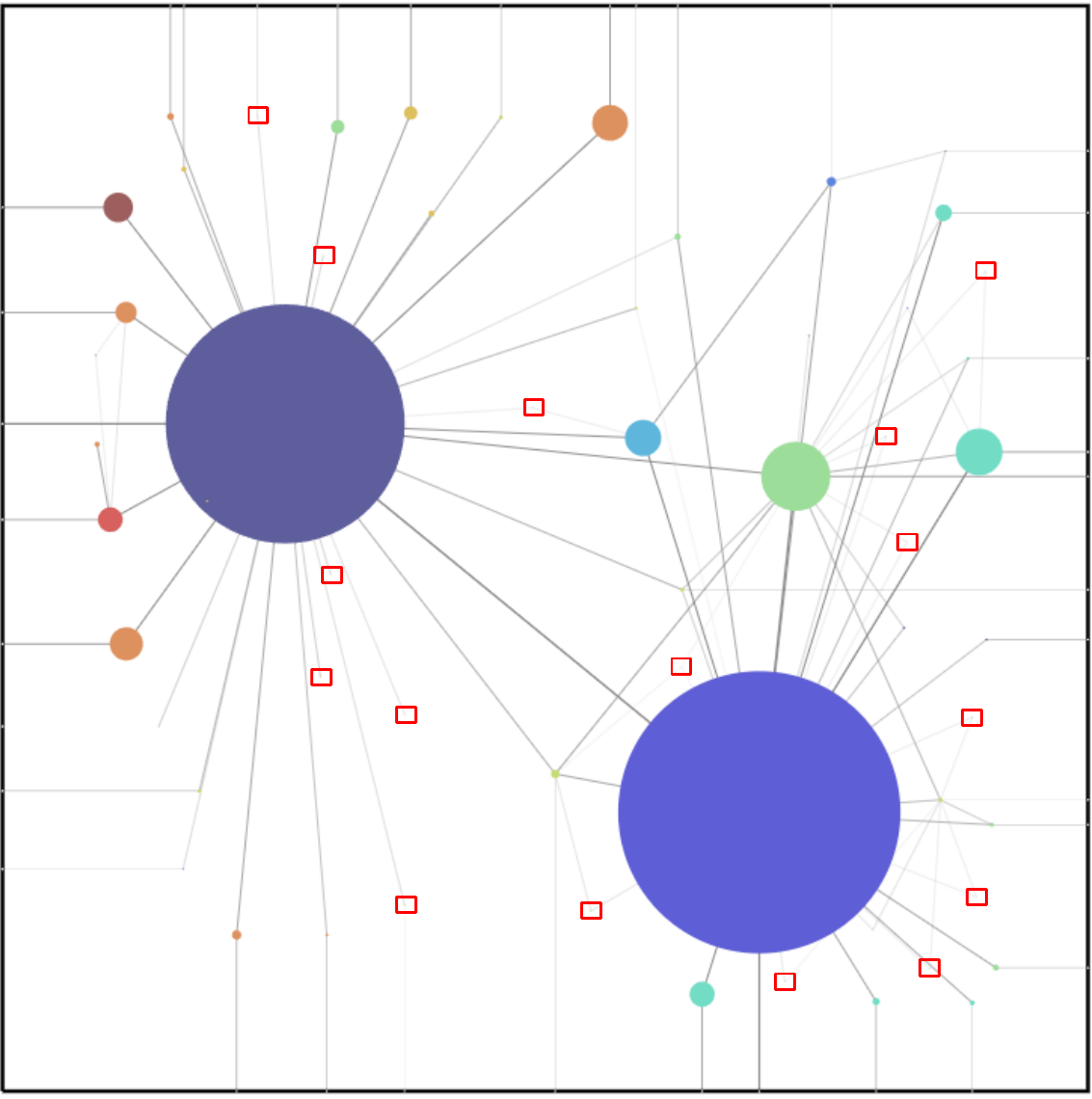}
         \caption{Node-link diagram}
         \label{fig:ablationGraph}
     \end{subfigure}
        \caption{Different visualizations of 3D medical image segmentation data. The volume visualization (a) only shows selected segments (here vessel, needle, and tumor). The 2D embedding (b) provides an overview of sizes and topological information of all segments. The given organ tags are used to label the embedding's segments. (c) Encoding the same information in a node-link diagram using the same color map as in (b). The red boxes mark very small nodes.}
        \label{fig:ablation}
\end{figure*}

\begin{figure*}[hbt]
     \centering
     \begin{subfigure}[b]{0.33\textwidth}
         \centering
         \includegraphics[width=\textwidth, alt={Three different node-link diagrams. Not using logarithmic scaling leads to large differences in sizes and smaller structures no longer being visible. Scaling the edges or edges and nodes logarithmically makes size comparisons difficult.}]{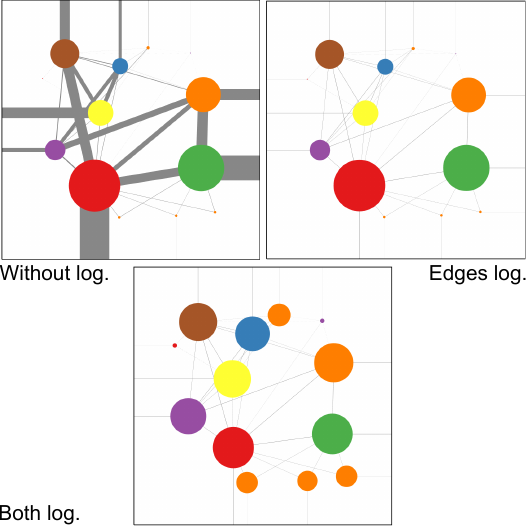}
         \caption{Node-link diagrams}
         \label{fig:semiconductorGraph}
     \end{subfigure}
     \begin{subfigure}[b]{0.33\textwidth}
         \centering
         \includegraphics[width=\textwidth, alt={Segmentation embedding for the parameter space partitioning of the semiconductor dataset.}]{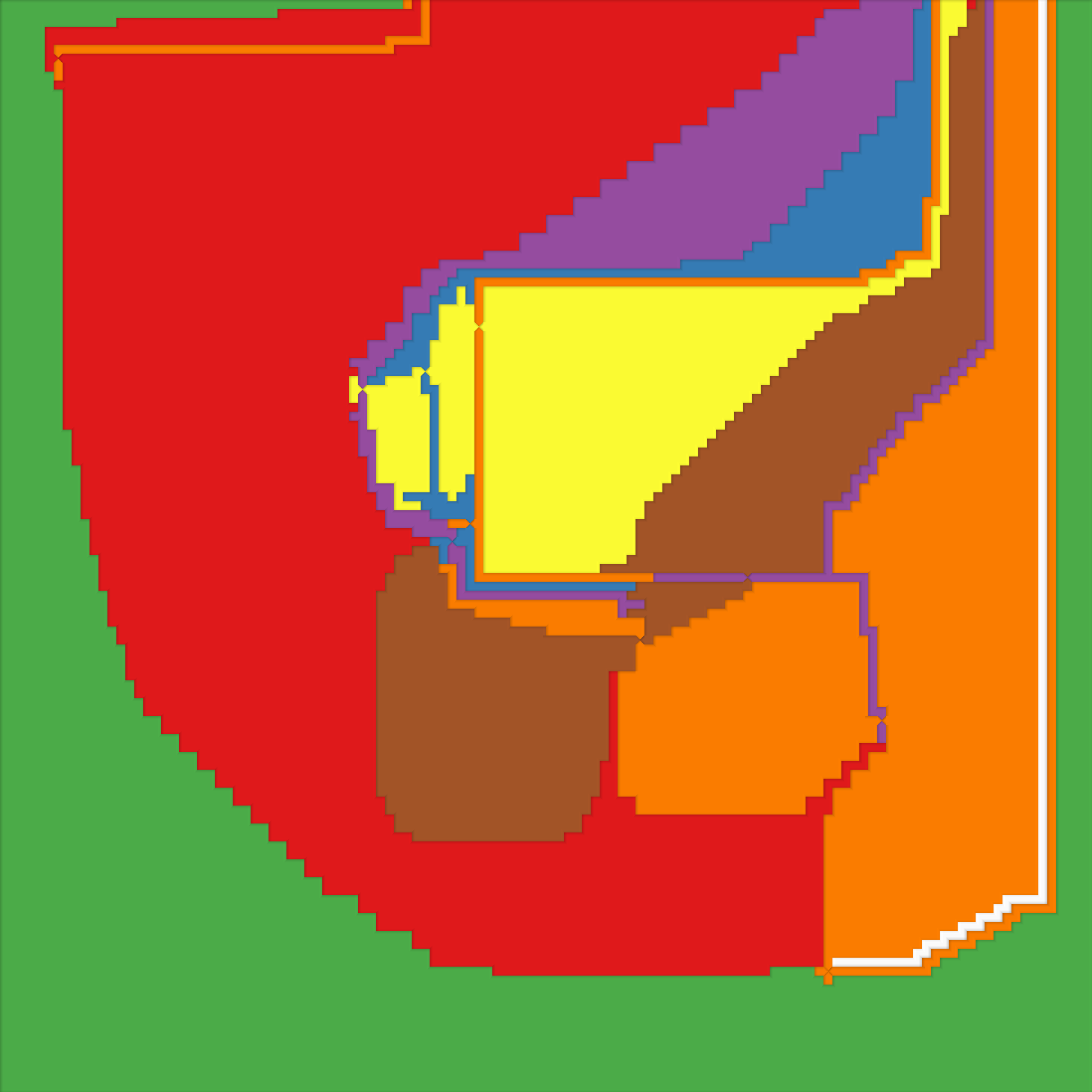}
         \caption{Segmentation embedding}
         \label{fig:semiconductorEmbedding}
     \end{subfigure}
     \begin{subfigure}[b]{0.33\textwidth}
         \centering
         \includegraphics[width=\textwidth, alt={Hyperslicer-visualization of the semiconductor dataset.}]{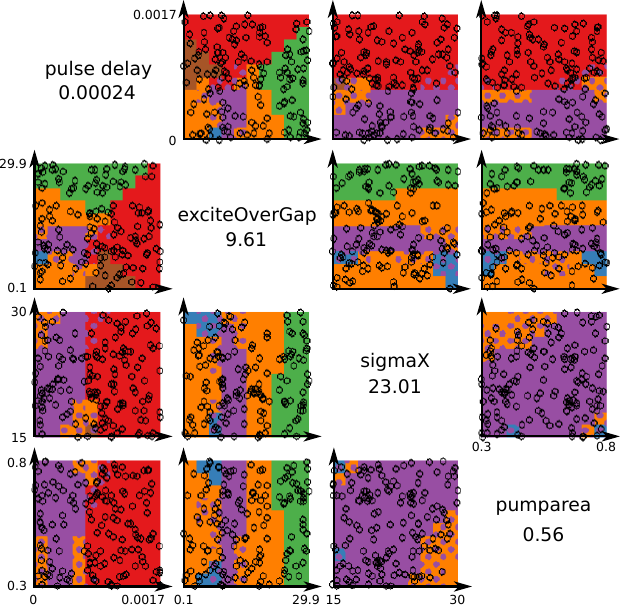}
         \caption{Hyper-slicer}
         \label{fig:semiconductorHyperslicer}
     \end{subfigure}
        \caption{Different encodings of the node-link diagram (a), 2D embedding (b) and hyper-slicer (c) of a 4D parameter space partitioning of a semiconductor simulation ensemble. The embedding allows the user to investigate the structures globally, while the hyper-slicer only provides a local view.}
        \label{fig:semiconductor}
\end{figure*}

\subsection{3D Image Segmentations}
\label{sec:3DResults}

As a first use case, we apply our algorithm to a 3D volume segmentation used as an input for a radiofrequency ablation simulation. The 3D domain covers the human body around the liver and contains different organ tags as well as tags for a tumor and the needle that is used for radiofrequency ablation. The dataset has a resolution of $92\times 92\times 92$ and contains $11$ different segments. As the regions with unique organ tags are not all connected, we set a unique ID for each connected component, which leads to $59$ segments. A volume visualization showing only a subset of the tissue types is shown in~\cref{fig:ablationVolume}. While a 3D segmentation could also be visualized in a volume rendering, showing a growing number of segments quickly leads to occlusion. Without interaction, rendering $11$ segments at the same time is already challenging. However, we do not consider our visualization to be a replacement for common volume visualization techniques but instead propose to use the segmentation embedding to obtain an overview of all segments and investigate their sizes and topology. For example, it could be used in an interactive setting, where segments of interest can be selected in the embedding and shown in a 3D surface or volume visualization.

The resulting embedding is shown in \cref{fig:ablationEmbedding}. Here, we apply a color coding based on the original segments. When observing the embedding, we can directly see that the majority of the given domain is covered by the two segments that are labeled as {liver} and {other tissue} (tissue without specific organ tags). In the center, we see a group of segments that represent the vessels. The vessels have joint boundaries. While the hepatic artery (HA) and the portal vein (PV) are connected to the unlabeled tissue, the hepatic vein (HV) contains no connection to this kind of tissue. We also spot very small segments belonging to the hepatic artery (HA) that are isolated inside of a region with no organ tags. As artery parts that are neither connected to the border of the domain nor to other vessels are not plausible, this indicates either a mistake in the segmentation or an undersampling artifact. 
The needle used for ablating the tumor is surrounded only by liver tissue, unlabeled tissue, and tumor. Therefore, we can deduce that the placement of the needle does not damage any of the other tissue types.

\begin{figure}[hbt]
    \centering
    \includegraphics[width=\linewidth, alt={Results for the blood flow simulation parameter space. a) shows the hyper-slices where different segments are indicated by color. b) shows the sample points as circles after applying a dimensionality reduction. Big circles capture the mean of the cluster. c) shows the segmentation embedding with four different segments.}]{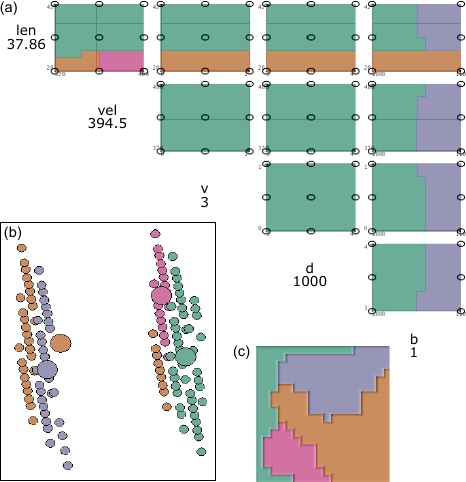}
    \caption{Parameter space visualizations of a 5D parameter space partitioning of a blood flow simulation ensemble, where the partitioning is based on a clustering of the ensemble members. Our segmentation embedding (c) provides a global overview that is not provided by the hyperslicer (a), yet the hyperslicer can provide some local details. The sample embedding (b) does not capture the topology of the multidimensional segmentation.}
        \label{fig:bloodflow}
\end{figure}

A straightforward alternative design choice for our visualization fulfilling all optimization criteria would be a representation of the segmentation properties in a node-link diagram as shown in \cref{fig:ablationGraph}. Details on how we create the node-link diagram can be found in the supplementary material, Section~\ref{sec:nodeLink}. The topological information can be encoded by the edges of the nodes, the sizes of the segments by the nodes's sizes and the boundary lengths by the edge width. Therefore, we explicitly compare such a node-link diagram to our segmentation embedding. 
As the boundary sizes vary significantly, we had to choose a logarithmic scale for the edge widths to keep most of the edges in a visible range. A segment's joint boundary with the border of the domain is encoded by an edge connecting to the black bounding box.

Node-link diagrams have been shown to not scale well with the number of nodes and an increasing node-link density~\cite{munzner2014visualization}. We also observe this issue when comparing our embedding to the node-link diagram. In general, the dense pixel visualization of our embedding makes better use of the available screen space. This becomes especially clear when encoding the segment sizes and edges. Some of the nodes representing smaller segments are barely visible such as the ones marked in \cref{fig:ablationGraph}. Additionally, the logarithmically scaled edge widths are hard to interpret with respect to what portion of a segment's boundary is shared with another. 
This problem is also shown in \cref{fig:semiconductorGraph}, where we compare the different options for a less complex node-link diagram. While the logarithmic scaling improves the visibility of edges and nodes such as the small red one, sizes are much harder to interpret and compare. For example, the size difference between the yellow and the large red node is hard to estimate (Tasks T2 and T3). In this example, we also observe occlusion of the yellow node and some edges, which complicates the understanding of the topological structure (T1). As the segmentation embedding resembles segmentation visualizations, segments completely enclosed by other segments in the high-dimensional domain are visualized as completely enclosed segments in the embedding, which is quite intuitive. In the graph drawing, instead, such segments are visualized as nodes with a single edge, where its interpretation requires an additional cognitive effort.

\subsection{Multi-dimensional Parameter Space Partitioning}
\label{sec:parameterSpace}

In this section, we analyze multi-dimensional parameter space partitionings of simulation ensembles. We first consider the analysis of a blood flow simulation ensemble with a 5D parameter space. This simulation ensemble was created to study the blood flow through a brain aneurysm and find biomarkers for certain diseases~\cite{leistikow2020interactive}. The simulation is driven by five different input parameters: a characteristic length, a characteristic velocity, the viscosity of the fluid, its density, and a parameter defining the boundary conditions. The simulation output can be used to partition the parameter space into regions leading to similar behavior. In this paper, we work on the parameter space partitioning as determined by Evers et al.~\cite{evers2022multi}. They presented an approach for partitioning the parameter space and analyzing it using a hyper-slicer as shown in \cref{fig:bloodflow}a. Their analysis approach also contains a parameter space overview visualization for which they apply multidimensional scaling to the parameter space samples which are the parameter settings for the simulation runs. Then, they color code the points in low-dimensional space based on the cluster they belong to and additionally insert a larger point representing the cluster for navigation (see~\cref{fig:bloodflow}b). 
However, this visualization does not support the investigation of segment sizes and topological properties. 
For the parameter space sampling, we choose a resolution of $10$ in each dimension. Our 2D embedding of the parameter space partitioning (\cref{fig:bloodflow}c) together with the parameter space visualizations proposed by Evers and Linsen~\cite{evers2022multi} is shown in \cref{fig:bloodflow}. The 2D embedding obtained a mean area deviation of $0.62\%$ and a mean boundary deviation of $0.38\%$, which indicates a very accurate representation. Our 2D embedding reveals that the green segment is connected to all other segments. This can be confirmed when viewing those 2D slices where we can see the shared boundaries. However, the 2D slices only show a subset of information. For example, we cannot immediately see that the largest part of the parameter space belongs to the orange segment or that the purple and the pink segments do not share a joint boundary. The parameter sample embedding shown in \cref{fig:bloodflow}b also does not allow us to obtain this information. For this use case, it does not become clear from the parameter sample embedding that the green and the purple segments share a joint boundary in parameter space. Thus, the topology information is not pertained. Additionally, the sizes could only be represented by the number of points belonging to each segment. However, if several points are placed close to each other in the parameter sample embedding, the size information is difficult to obtain. Hence, we conclude that our 2D segmentation embedding allows for a better global overview of the entire structure of the multi-dimensional partitioning.

Next, we analyze the 4D parameter space partitioning for a semiconductor simulation ensemble~\cite{evers2022multi}. The simulation ensemble was created to study transport properties in a quantum wire. As in the previous use case, the partitioning was created by clustering the ensemble members and transferring this clustering to the 4-dimensional parameter space of the simulation. We choose the same clustering as Evers and Linsen~\cite{evers2022multi}, which results in 7 clusters. The parameter space is sampled with a resolution of $10$ in each dimension. Our 2D embedding of the parameter space partitioning is shown in \cref{fig:semiconductorEmbedding}, where we color the segments according to the clusters. The embedding shows that the parameter space partitions of the clusters are not necessarily connected. For example, the red and purple segments are split into two disconnected regions. The same partitioning is shown in the hyper-slicer in \cref{fig:semiconductorHyperslicer}. Obtaining the same structural information would not only require a significant amount of interaction with the hyper-slicer but also a high amount of mental effort.

\section{Discussion and Conclusion}
We presented an algorithm for embedding multi-dimensional partitionings into 2D,  where we optimize the area and boundary sizes to represent those of the multi-dimensional space while preserving the partitioning's topology. The 2D embedding was computed by first creating a graph representation of the partitioning, whose graph drawing was used as an initial configuration for cellular automaton optimization. Our rendering of the embedded segments allows for a large flexibility in the choice of color codings. 

Our approach maintains the desired criteria well and allows us to represent several structures that could occur in segmentations including segments that are completely surrounded by other segments and connections to the boundary. Besides an evaluation on synthetic datasets, an application to 3D segmentations as well as to multi-dimensional parameter space partitionings shows its utility on real-world data. In particular, we demonstrated that our approach provides an overview of the entire segmentation structure, which other approaches, such as a hyper-slicer or dimensionality reductions, cannot provide. We also compared the segmentation embedding to representations based on a node-link diagram and showed that our approach presents certain features of segmentations, such as enclosed segments, more clearly and uses the available space more efficiently. Interpreting our visualization requires no knowledge about graphs as the underlying graph representation is not visible in the final visualization. Additionally, we showed scaling problems of the node-link diagram. However, a deeper investigation of the strengths and weaknesses of both approaches would require a user study, which we plan for future work. 

Our embedding provides an overview of the segmentation structure but loses information about the positions and shapes of the segments. However, linking our 2D embedding to coordinated views of volume renderings of 3D segments or multi-dimensional data visualizations (like the hyper-slicer for parameter space partitionings or parallel coordinates) allows for accessing this information for selected segments, where the selection can easily be performed in our 2D embedding. A limitation of the current implementation are relatively high computation times. However, the implementation is not yet optimized and parallelized, which could lead to a significant speed-up. On the other hand, the computation of the 2D embedding can be a pre-processing step for an interactive analysis and only needs to be computed once. If only small changes in the sizes of the segments and boundaries occur, it would also be possible to update the embedding instead of requiring a complete recomputation. Thin segments might occur in order to preserve the topology and in the attempt to optimize the boundary, which may hamper area estimation. We plan to address this limitation by adapting the cellular automaton in future work. Incorporating our 2D embedding into more powerful interactive analyses such as analyzing the simulation outcomes of a parameter-space segment is another potential future research direction, but was clearly beyond the scope of this paper. Moreover, the quality in preserving the boundary length is lower than that of the area preservation, which provides room for further improvements.

\section*{Supplemental Materials}
\label{sec:supplemental_materials}
The implementation of the algorithm is available at \url{https://github.com/marinaevers/segmentation-projection}. The supplemental material published together with this paper contains an appendix with further details for reproducibility and a video that shows the iterative optimization of the algorithm.

\section*{Acknowledgments}
This work was funded by the Deutsche Forschungsgemeinschaft (DFG, German Research Foundation) grant 260446826 (LI 1530/21-2).

\bibliographystyle{abbrv-doi-hyperref}

\bibliography{refs}

\clearpage

\appendix 

\renewcommand{\thefigure}{A\arabic{figure}}

\setcounter{figure}{0}

\begin{figure*}[hbt]
     \centering
     \begin{subfigure}[b]{0.4\textwidth}
         \centering
         \includegraphics[width=\textwidth]{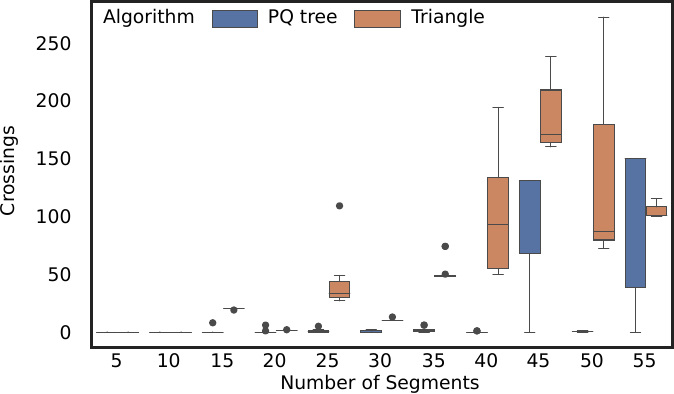}
         \caption{2D}
         \label{fig:sub2D}
     \end{subfigure}
     \begin{subfigure}[b]{0.4\textwidth}
         \centering
         \includegraphics[width=\textwidth]{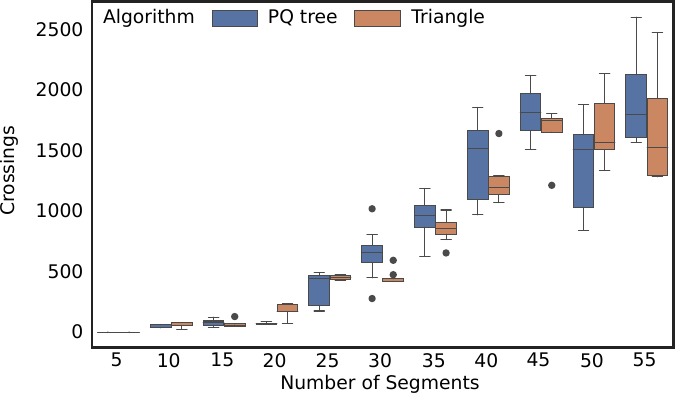}
         \caption{3D}
         \label{fig:sub3D}
     \end{subfigure}
        \caption{The number of edge crossings of the graph embedding could vary widely depending on the choice of the outer face. (a) For planar 2D graphs, the PQ-tree-based algorithm introduced no edge crossings in contrast to the triangle-based algorithm. (b) For 3D graphs, there is no clear winner. Using the minimum number of crossings could generally improve the results substantially.}
        \label{fig:crossings}
\end{figure*}

\section{Details of Graph Embedding and Initial Configuration}
In the following, we will present the detailed considerations to obtain a graph embedding that minimizes edge crossings and places the boundary vertex $v_b$ at the outer boundary of the graph. After that, we describe the implementation details to obtain an initial condition for the cellular automaton.

\subsection{Graph Embedding}
While force-based graph drawing approaches have been shown experimentally to keep the number of edge crossings comparatively small, we decided against using them, as they do not minimize edge crossings explicitly, which might lead to edge crossings even in planar graphs. Also, adding vertices like the one representing the domain's border imposes further constraints on the layout algorithm.
Orthogonal graph layouts, on the other hand, can be easily adapted to deal with the border vertex. Moreover, the purely horizontal and vertical edges facilitate the graph rasterization for the input of the cellular automaton and avoid many special cases that could arise, especially with diagonal edge crossings. As we want to use the embedding as a starting point for the cellular automaton approach that ensures topology preservation, we will remove the border vertex after the embedding and replace it with connections to the actual image border. 

For the graph embedding, we use the Open Graph Drawing Framework (ogdf)~\cite{chimani2007open}, which provides a customizable implementation of the topology-shape-metrics approach. We choose a topology-shape-metrics approach that minimizes the number of edge crossings in various steps~\cite{tamassia1988automatic}.
The basic idea is to first compute a maximum planar subgraph, which is then used for creating a planarization of the graph that does not exhibit any edge crossings but replaces them by dummy vertices. This planarization can be embedded and leads to a planar layout. Next, the edge crossings are reinserted and the positions of the nodes are computed.
As we use existing algorithms here, we will not discuss the details of the algorithm but focus on why it is suitable for our requirements. As we will discuss in the supplementary material,
the quality of the initial graph drawing influences the final result strongly. Hence, careful considerations are necessary.

Unfortunately, creating a maximum planar subgraph is also an NP-hard problem~\cite{yannakakis1978node}. Luckily, different approximation algorithms exist. We tested the triangle-based algorithm by Chalermsook and Schmid~\cite{chalermsook2017finding}, which was also used by Chimani et al. ~\cite{chimani2021star}, and a planarization algorithm using PQ-trees~\cite{Jayakumar1989PQ}, which is the default option of the ogdf framework. Their performance is evaluated in \cref{sec:crossings}.
Starting from the (approximated) maximum planar subgraph, additional edges are inserted for non-planar graphs. In this step, dummy vertices are inserted for edge crossings, thus, leading again to a planarization of the graph. Two techniques that can be applied here are the insertion of single edges and the star insertion, where vertices (including the incident edges) are reinserted.
A recent study showed that a mixed insertion planarizer performs best among all planarization algorithms for crossing minimizations in graph drawing~\cite{chimani2021star}.
Therefore, we decided to also use this method for our approach. The outcome is a planarized graph with dummy vertices.

In the next step, we need to define an embedding that places the vertex $v_b$ representing the boundary on the external face, where a face of a graph is a region bound by edges. Each face is identified uniquely by the order of its edges. The external face of a graph is the face that is only bound by edges inwards and thus corresponds to the outside of the graph drawing in a 2D plane. By computing a Boyer-Myrvold planar embedding~\cite{boyer2006simplified} of the planarized graph, we can define a combinatorial embedding (embedding defined by the order of edges for each vertex without explicit vertex positions and without external face). To ensure the border vertex $v_b$ to be located on the external face, we choose one of the faces that contains the border vertex $v_b$ as the external face. Note that the number of faces containing the vertex $v_b$ is equal to the vertex's degree. Thus, different embeddings that fulfill the criterion above are possible of which we choose the one that leads to the smallest number of edge crossings.

After embedding the planarized graph, we can apply an orthogonal graph layout algorithm~\cite{chimani2013open, tamassia1987embedding} to obtain a graph drawing that uses vertical and horizontal lines only. In this last step, the dummy vertices are replaced by edge crossings. 

\subsection{Initial Configuration.}
First, we need to determine a suitable grid resolution for the cellular automaton based on the extent of the graph embedding created in the previous step. Here, we assume the vertices to have a size of $20\times20$ cells and a minimum separation of $20$ cells. To ensure enough space for all edges and at least one separating cell between the different edges, we apply a scaling factor $f$ of the resolution. Since the border vertex $v_b$ might have a particularly high degree, we heuristically set the scaling factor to $f=\max(2, \sqrt{\mathrm{deg}(v_b)})$, where $\mathrm{deg}(v_b)$ denotes the degree of the border vertex. As the number of cells scales quadratically with this factor, the square root is used to ensure that the factor is growing slowly, which avoids a lack of main memory.

For assigning cells to graph edges, we can apply a naive line drawing algorithm, as no diagonal edges exist in an orthogonal graph drawing. Each edge is drawn using a line width of one cell. The ogdf-framework used for the graph drawing also provides the bending points of the edges, which we can directly use. We first draw all edges that do not involve the border vertex $v_b$.
Given an edge $(v_i,v_j)$, we traverse the corresponding line segments, where the first half of the traversed line segments are assigned the ID of vertex $v_i$ and the second half the ID of vertex $v_j$, while a central line segment is split between both segments.
When edge crossings occur, we label the cell with ID -2, which will allow us to preserve the topology during the iterative optimization (see below). 
Then, we draw all vertices except for border vertex $v_b$ by filling a square of cells with the vertex' ID, see \cref{fig:overview} for an example, where background cells are shown in grey and edge crossing cells in black.

For each vertex $v_i$ that contains a connection $(v_i,v_b)$ to the border, we would draw the edge $(v_i,v_b)$  from $v_i$ to $v_b$ and then continue by drawing a connection from $v_b$ to the border of the cellular automaton's grid. Remember that $v_b$ was placed close to the border. 
To account for potential overlaps of the edge drawings, we add an offset to the last bending point that varies between the edges that need to be connected to the border.
In case of obstacles that cannot be passed by an edge crossing, like vertices or other edges going in the same direction, we apply a re-routing by introducing an offset of $2$ cells until the edge's destination can be reached. 
If vertex $v_i$ itself was already close to the grid's border, we can use a simpler solution and directly draw an edge from $v_i$ to the closest point on the grid border, i.e., without going through $v_b$.
More precisely, if $v_i$ is closer to the nearest border than two times the size of the border vertex, we can assume that no other vertex is between $v_i$ and the border and draw the direct connection to the border.

As described above, for the graph embedding, we set the external face such that it contains the border vertex. We mentioned that multiple choices exist and that we want to select the one where edge crossings are minimized. However, the simple edge routing used here to compute the initial configuration of the cellular automaton might introduce additional edge crossings depending on the exact edge routes. 
Thus, we compute different graph embeddings and transform them to the initial configurations and then choose that graph embedding where the number of edge crossings is minimal in the initial configuration. 
This procedure also increases our algorithm's stability because, in some cases, a topology-preserving transformation might not be possible with our edge routing algorithm. In this case, we neglect this attempt and choose the initial configuration with the minimum edge crossings of the remaining attempts.

After finding a suitable initial configuration for the cellular automaton, we reduce the number of cells to reduce computation times. For this step, we remove each row and column where at least one neighboring row or column contains exactly the same cells. Using this procedure, we only reduce the sizes of nodes and distances (see the graph drawing in \cref{fig:overview} of the main article), without violating the topology.

\section{Graph Embedding's Influence on Edge Crossings.}
\label{sec:crossings}
Our algorithm's output quality strongly depends on the initial graph embedding. Even though the graph embedding algorithm is exchangeable and we do not want to focus too much on graph drawing here, we want to evaluate how much the number of edge crossings varies only by choice of the graph embedding. We use dataset $D_1$ with varying numbers of segments from $5$ to $55$ in 2D and 3D settings. Additionally, we want to study the influence of the choice of the external face in the combinatorial embedding. We compute the number of edge crossings for all different choices of external faces, where we only consider successful embeddings. Here, we compare the default PQ tree-based algorithm to the triangle-based algorithm. The results for the number of edge crossings are shown in \cref{fig:crossings}. In the 2D case, the graph is obviously planar and we observe that the PQ tree-based algorithm always allows finding embeddings without edge crossings, while this is not the case for the triangle-based algorithm (see \cref{fig:sub2D}). In general, the number of edge crossings spans a wide range. Thus, choosing the embedding with the lowest number of edge crossings allows for significant improvements. Comparing both subgraph computation algorithms in the 3D case (see \cref{fig:sub3D}), we observe that the triangle-based algorithm provides better results in some cases (e.g., for $55$ segments), while being worse in other cases (e.g., for $50$ segments). In all cases, the number of edge crossings is in the same order of magnitude. Thus, we cannot clearly recommend which algorithm to use. However, since the PQ tree-based algorithm performed better for planar graphs, we opt for using it for all examples presented in this article. However, the algorithm could be easily exchanged, if prior knowledge of the segmentation to embed indicates that another algorithm is more suitable.

In some cases, the number of crossings could decrease further by choosing a more sophisticated edge re-routing algorithm for the edges towards the boundary. Our method of testing different graph embeddings and choosing the one with minimum edge crossings yields sufficiently good results though. 

\section{Evaluation}
\label{sec:appendixEvaluation}
All timings reported in this section were obtained on a laptop with a 1.7GHz AMD Ryzen Pro 7 processor.
\begin{table}[hbt]
\centering
\caption{Quality criteria for different datasets with dimension $n$ and number of segments $|S|$. Crossings denote the number of edge crossings after finishing the embedding, and resolution provides the image resolution after finishing the cellular automaton.}
\resizebox{\linewidth}{!}{%
\begin{tabular}{|l|c|c|c|c|c|c|c|}
\hline
\textbf{Dataset} & \textbf{$n$} & \textbf{$|S|$} & \textbf{Crossings} & \textbf{Resolution} & \textbf{$\bar{d_A}$} (\%) & \textbf{$\bar{d_L}$} (\%) \\
  \hline  \hline
  $D_1$ & 2 & 20 & 0 & $116\times 134$ & 0.017 & 0.963 \\ \hline
  $D_2$ & 3 & 8 & 4 & $46\times 28$ & 0.089 & 5.517 \\ \hline
  Ablation & 3 & 59 & 14 & $374\times 448$ & 0.547 & 1.708\\ \hline
  Nucleon & 3 & 42 & 17 & $208\times 270$ & 3.138 & 1.766 \\ \hline
  Blood flow & 5 & 4 & 0 & $18\times 18$ & 0.170 & 0.495 \\ \hline
  Semiconductor & 4 & 13 & 13 & $114\times 128$ & 1.576 & 4.045 \\ \hline
\end{tabular}}
\label{tab:embeddingDatasets}
\end{table}

\begin{figure}[htb]
     \centering
     \includegraphics[width=\linewidth]{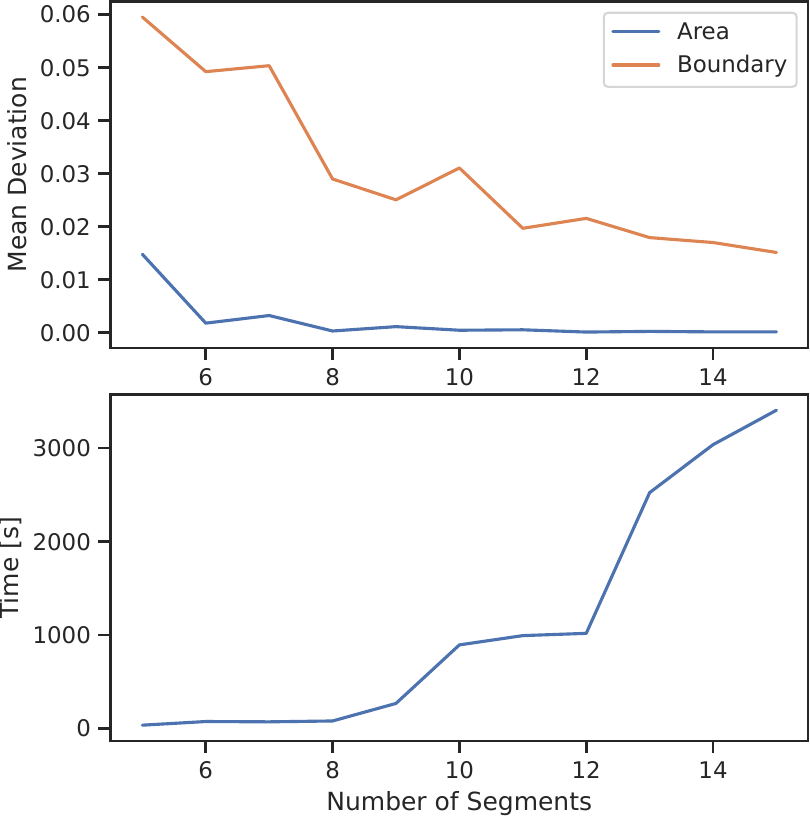}
    \caption{While the deviation in area and boundary length decreases with an increase in the number of segments, the increased graph complexity leads to an increase in computation time.}
    \label{fig:scaling}
\end{figure}

\begin{table}[hbt]
\centering
\caption{Timings for the individual steps of different datasets with $N$ input voxels and $M$ pixels in the output.}
\resizebox{\linewidth}{!}{%
\begin{tabular}{|l|c|c|c|c|c|c|}
\hline
\textbf{Dataset} & $N$ & $M$ & \textbf{Graph (s)} & \textbf{Embedding (s)}  & \textbf{Automaton (s)} \\
  \hline  \hline
  $D_1$ & 2,500 & 15,544 & 0.006 & 0.576 & 244.122 \\ \hline
  $D_2$ & 8,000 & 1,568 & 0.037 & 0.353 & 45.706 \\ \hline
  Ablation & 778,688 & 167,552 & 2.851 & 129.784 & 4809.0 \\ \hline 
  Nucleon & 68,921 & 56,160 & 0.245 & 1.669 & 1112.2 \\ \hline
  Blood flow & 3,200,000 & 324 & 26.446 & 0.009 & 3.294 \\ \hline
  Semiconductor & 10,000 & 14,592 & 0.090 & 5.562 & 257.648 \\ \hline
\end{tabular}}
\label{tab:timingsDatasets}
\end{table}

\begin{figure}[htb]
     \centering
     \includegraphics[width=\linewidth]{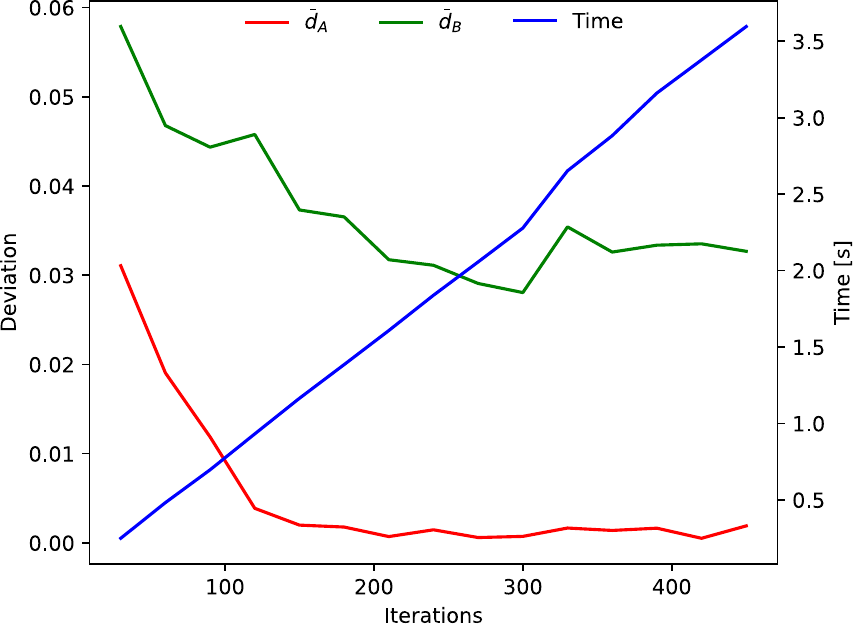}
    \caption{While the computation time increases approximately linearly, the quality metrics initially decrease quickly and stop decreasing for larger numbers of iterations.}
    \label{fig:iterations}
\end{figure}

To investigate the influence of the number of iterations, we study the computation time and the output quality for a varying number of iterations. We use dataset $D_1$ with three dimensions and eight segments. For the timings, we only consider the evaluation of the cellular automaton, as the computation of the graph, the computation of the embedding, and the time for the rendering are independent of the number of iterations. The results are shown in \cref{fig:iterations}. As expected, the computation time increases approximately linearly while the additional costs for removing unnecessary segments become visible at $300$ iterations. The quality metrics show a clear decrease in the beginning before fluctuating around a certain level. Therefore, we can conclude that early stopping is justified.

\section{Graph Visualization for Comparison}
\label{sec:nodeLink}
For creating the node-link diagrams, we use a force-based graph drawing algorithm based on the Fruchterman-Reingold algorithm~\cite{fruchterman1991graph}. However, we adapt the algorithm to improve the final layout to meet our needs. For the distances between the vertices, we take the area we will need for the vertices into account. Thus, we replace the constant $k$ that denotes the optimal distance between vertices in the original algorithm by a value $k'_{ij}=k_0+r_i+r_j$ at the edge between vertices $v_i$ and $v_j$ where $r_i$ and $r_j$ denote the radii of the final circles that will represent the vertices. The parameter $k_0$ represents an additional offset between the vertices and is chosen as $k_0=\sqrt{1/|S|}$. For each vertex $v_i$ connected to the boundary, we create a second vertex $v_{-i}$ that always stays at the boundary of the unit cube but is moved to the closest point on the boundary. We also limit the placement of the vertices to the unit square area. To prevent vertices from leaving this area, the boundary exerts a repulsive force on vertex $v_i$ which is computed as $F_{b,i}=(r_i+k)^2*(1/x^2-1/(1-x)^2, 1/y^2-1/(1-y)^2)$ on the nodes, where $x$ and $y$ are the coordinates of the vertex. By squaring the distances, the force decreases quickly with increasing distance. For both node-link diagrams presented in this paper, we use 1000 iterations, a step size of 0.001, and normalize the area sizes such that the radius of the largest vertex is 0.1. We also apply simulated annealing to aid convergence and reduce the step size by 1\% in each step. For the edges, we apply a logarithmic scaling to the sizes to avoid several edges vanishing because of their small size.

\section{Use Case: 3D Nucleon Segmentation}
\begin{figure*}[htb]
     \centering
     \begin{subfigure}[b]{0.2\textwidth}
         \centering
         \includegraphics[width=\textwidth]{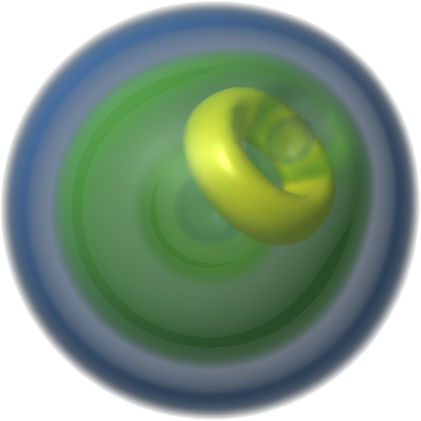}
         \caption{Volume visualization}
         \label{fig:nucleon-volume}
     \end{subfigure}
     \hfill
     \begin{subfigure}[b]{0.46\textwidth}
         \centering
         \includegraphics[width=\textwidth]{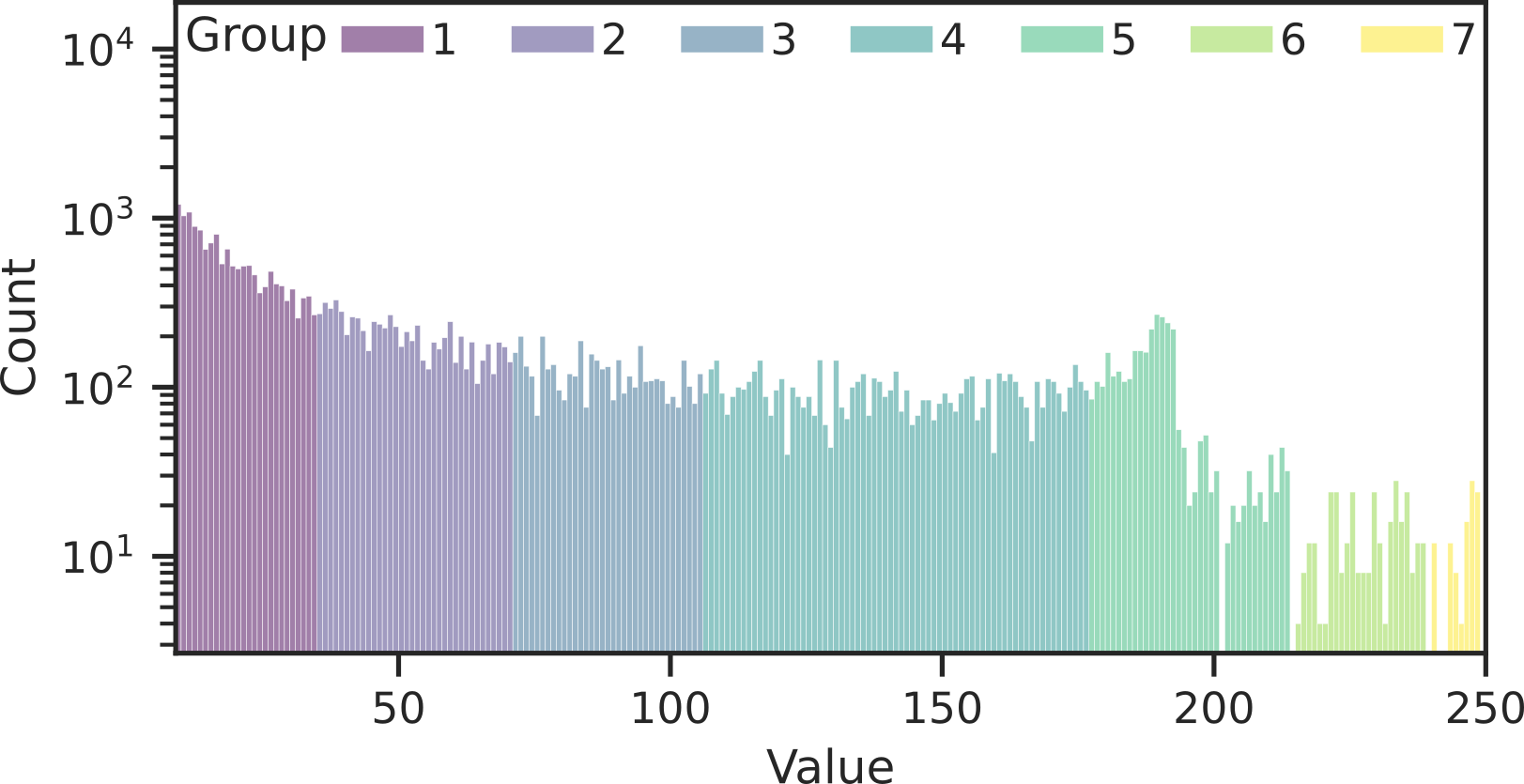}
         \caption{Histogram}
         \label{fig:nucleon-histogram}
     \end{subfigure}
     \hfill
     \begin{subfigure}[b]{0.33\textwidth}
         \centering
         \includegraphics[width=\textwidth]{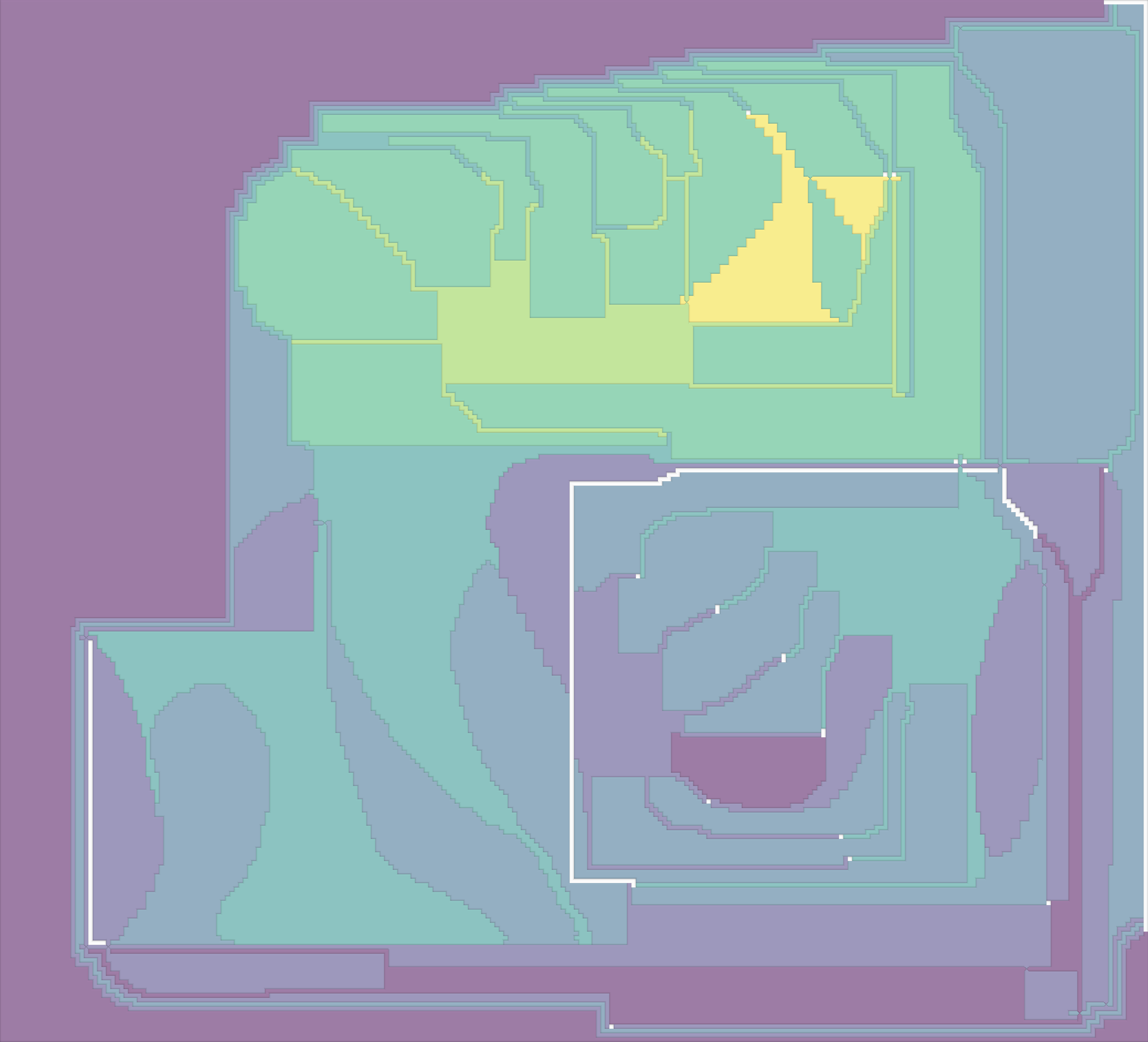}
         \caption{Embedding}
         \label{fig:nucleon-embedding}
     \end{subfigure}
    \caption{The nucleon dataset is partitioned using 7 bins of the histogram leading to 42 segments. In the 2D embedding, all segments belonging to the same bin are encoded by the same color. The volume visualization only shows three selected segments, while the 2D embedding allows for showing all segments simultaneously.}
    \label{fig:nucleon}
\end{figure*}
The nucleon dataset is a 3D volume that describes the two-body distribution probability of a nucleon based on a simulation. We partition the nucleon dataset shown in \cref{fig:nucleon-volume} based on the histogram as shown in  \cref{fig:nucleon-histogram}. We identify connected components, which results in $42$ segments. The 2D embedding is shown in \cref{fig:nucleon-embedding}. An animation of the iterative computation is shown in the supplementary material. We used $2,000$ iterations and obtained a mean area deviation of $2.4\%$ and a mean boundary deviation of $1.5\%$. Here, we assigned the same color to all segments that belong to the same range in the histogram. In this way, we can investigate the structure of the segmentation. For example, we can observe that values of the histogram region encoded in dark blue appear not only as surrounding the nucleon but also in two regions in the inside.  We also see that the yellow and light green region are in general surrounded by the region labeled as $5$ in \cref{fig:nucleon-histogram}, but the segments also connect to the region labeled as $6$. In general, we can observe that the histogram regions $3$ and $5$ are split into several individual segments.

\begin{figure}[htb]
     \centering
     \includegraphics[width=\linewidth]{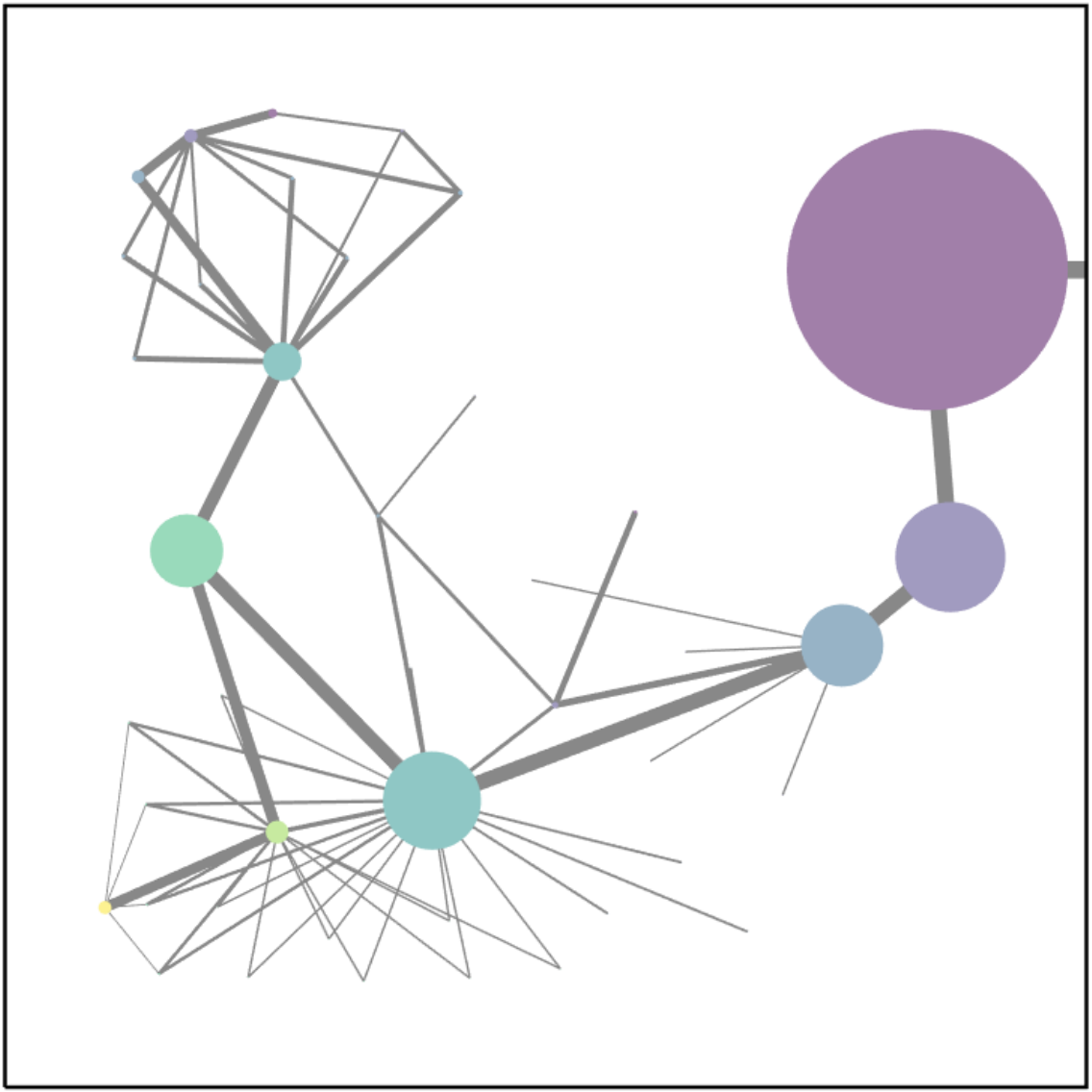}
    \caption{Node-link diagram for the nucleon dataset. The color coding is consistent with \cref{fig:nucleon}. The edges encode the boundary lengths on a logarithmic scale and the node sizes segment sizes. Some nodes vanish due to the small scale.}
    \label{fig:nucleonNodeLink}
\end{figure}

For comparison, we also show the node-link diagram of this dataset in \cref{fig:nucleonNodeLink}. When comparing this visualization, we basically observe the same results as for the node-link diagram of the ablation dataset. Even though the number of segments is lower in this example, the visualization is already very complex and structure is harder to interpret than for the embedding.

\end{document}